\documentclass[10pt,oneside,onecolumn,floatfix]{article}
\usepackage{epsfig}
\usepackage{a4wide}
\usepackage[top=30pt,bottom=30pt,left=48pt,right=46pt]{geometry}
\usepackage{slashed, enumitem}
\usepackage[utf8]{inputenc}
\usepackage{jcappub}
\usepackage{float,color}
\usepackage{subfigure}
\usepackage{multirow}
\usepackage{slashed}
\usepackage{placeins}
\usepackage{wasysym} 
\usepackage{mathrsfs} 
\usepackage{caption}
\usepackage{amsmath}
\usepackage{amssymb}
\usepackage{graphicx}
\usepackage{slashed}
\usepackage{multirow}
\usepackage{placeins}
\usepackage[dvipsnames]{xcolor}
\usepackage{epstopdf}
\usepackage{soul}
\usepackage{tikz}
\usepackage[capitalise, english]{cleveref}
\usepackage{siunitx}
\usepackage{xspace}
\usetikzlibrary{trees}
\usetikzlibrary{decorations.pathmorphing}
\usetikzlibrary{decorations.markings}
\usepackage[colorlinks=true,citecolor=blue,linkcolor=blue]{hyperref}

\usepackage{mhchem}

\usepackage{subfigure}

\usepackage{upgreek}
\usepackage{mathtools}
\usepackage{setspace, slashed, gensymb, hhline, booktabs}
\usepackage{wrapfig}
\usepackage{makecell}

\usepackage{pifont}

\definecolor{viridian}{rgb}{0.25, 0.51, 0.43}
\definecolor{mediumseagreen}{rgb}{0.24, 0.7, 0.44}
\definecolor{otterbrown}{rgb}{0.4, 0.26, 0.13}
\definecolor{saddlebrown}{rgb}{0.55, 0.27, 0.07}
\definecolor{americanrose}{rgb}{1.0, 0.01, 0.24}
\definecolor{ao}{rgb}{0.0, 0.0, 1.0}

\usepackage{xcolor}
\definecolor{notecolor}{rgb}{0.8. 0.0, 0.0}

\newcommand\myshade{80}
\colorlet{mylinkcolor}{violet}
\colorlet{mycitecolor}{red}
\colorlet{myurlcolor}{ao}

\hypersetup{
	linkcolor  = mylinkcolor!\myshade!black,
	citecolor  = mycitecolor!\myshade!black,
	urlcolor   = myurlcolor!\myshade!black,
	colorlinks = true
}

\newcommand{\ud}{\mathrm{d}}   
\newcommand{\ue}{\mathrm{e}}   
\usepackage{upgreek}

\usepackage[utf8]{inputenc}
\usepackage{siunitx}
\DeclareSIUnit\parsec{pc}
\usepackage[numbers]{natbib}
\usepackage{appendix}
\usepackage{xcolor}

\usepackage{tikz,xcolor,hyperref}

\definecolor{lime}{HTML}{A6CE39}
\DeclareRobustCommand{\orcidicon}{\hspace{-3mm}
	\begin{tikzpicture}
	\draw[lime, fill=lime] (0,0) 
	circle [radius=0.16] 
	node[white] {\hspace{0.1mm}{\fontfamily{qag}\selectfont \tiny ID}};
	\draw[white, fill=white] (-0.07,0.1) 
	circle [radius=0.01];
	\end{tikzpicture}
	\hspace{-5mm}
}

\foreach \x in {A, ..., Z}{\expandafter\xdef\csname orcid\x\endcsname{\noexpand\href{https://orcid.org/\csname orcidauthor\x\endcsname}
		{\noexpand\orcidicon}}
}

\title{Constraining primordial black holes as dark matter using the global 21-cm signal with X-ray heating and excess radio background} 
\author{Shikhar Mittal\orcidA{},}
\emailAdd{shikhar.mittal@tifr.res.in}
\author{Anupam Ray\orcidB{},}
\emailAdd{anupam.ray@theory.tifr.res.in}
\author{Girish Kulkarni\orcidC{},\\ and}
\emailAdd{kulkarni@theory.tifr.res.in}
\author{Basudeb Dasgupta\orcidD{}}
\emailAdd{bdasgupta@theory.tifr.res.in}
\affiliation{Tata Institute of Fundamental Research, Homi Bhabha Road, Mumbai 400005, India}

\abstract
{
Using the global 21-cm signal measurement by the EDGES collaboration, we derive constraints on the fraction of the dark matter that is in the form of primordial black holes (PBHs) with masses in the range $\num{e15}$--$\SI{e17}{\gram}$. Improving upon previous analyses, we consider the effect of the X-ray heating of the intergalactic medium on these constraints, and also use the full shape of the 21-cm absorption feature in our inference. In order to account for the anomalously deep absorption amplitude, we also consider an excess radio background motivated by LWA1 and ARCADE2 observations. Because the heating rate induced by PBH evaporation evolves slowly, the data favour a scenario in which PBH-induced heating is accompanied by X-ray heating. Also, for the same reason, using the full measurement across the EDGES observation band yields much stronger constraints on PBHs than just the redshift of absorption. We find that 21-cm observations exclude $f_{\mathrm{PBH}} \gtrsim \num{e-9.7}$ at 95\% CL for $M_{\mathrm{PBH}}=\SI{e15}{\gram}$. This limit weakens approximately as $M_{\mathrm{PBH}}^4$ towards higher masses, thus providing the strongest constraints on ultralight evaporating PBHs as dark matter over the entire mass range $\num{e15}$--$\SI{e17}{\gram}$. Under the assumption of a simple spherical gravitational collapse based on the Press-Schechter formalism, we also derive bounds on the curvature power spectrum at extremely small scales ($k\sim \SI{e15}{\mega\parsec^{-1}}$). This highlights the usefulness of global 21-cm measurements, including non-detections, across wide frequency bands for probing exotic physical processes.
}

\subheader{TIFR/TH/21-6}
\keywords{physics of the early universe; high redshift galaxies; primordial black holes; dark matter theory}

\begin{document}
\maketitle
\flushbottom
\section{Introduction}
\label{intro}

The 21-cm signal --- the brightness temperature of the $\SI{1420.4}{\mega\hertz}$ hyperfine transition of cosmological neutral hydrogen measured against the cosmic microwave background (CMB) temperature --- has been proposed as a probe of Cosmic Dawn and the Epoch of Reionization (EoR) \cite{MMR}. Observing this signal is a challenging task due to large foregrounds, mainly the galactic synchrotron radiation \cite{Peng, oliveira, bernardi, tegmark}. In 2018, the EDGES collaboration (Experiment to Detect the Global EoR Signal) \cite{edges} reported the first and till-date the only measurement of the cosmological 21-cm signal \cite{Bowman}. The key features of this absorption signal are its location $(\SI{78.2}{\mega\hertz}$, corresponding to $z=17.2$), amplitude $(\Delta T_{\mathrm{b}}\approx\SI{-500}{\milli\kelvin})$ and the full width at half maximum $(\SI{19}{\mega\hertz})$. The most interesting and intriguing part of their detected signal is the amplitude, which is found to be more than double the prediction of even the most optimistic theoretical models.

The spin temperature of the hyperfine energy states of neutral hydrogen (\textsc{H\,i}) is the main quantity that decides the amplitude of the 21-cm signal. At redshifts $z\lesssim30$, the physics of the spin temperature is expected to be relatively simple. At these redshifts, it is mainly affected by the standard physics of \textsc{H\,i} and Lyman-$\alpha$ (Ly~$\alpha$) interaction via the Wouthuysen-Field effect \cite{Wouth, Field}. Other relevant processes are the adiabatic cooling of the intergalactic medium (IGM) and small heating effects due the Ly~$\alpha$ radiation \cite{Chen, Meiksin, FP06, Mittal} and the ambient 21-cm background \cite{Venu}. The shape of the absorption signal measured by EDGES is potentially also decided by processes such as Compton heating \cite{Weymann}, X-ray heating \cite{F06, Mesinger_13, Christian_2013, Fialkov_17} and reionization \cite{Haard, Madau_2017}. The disagreement between models that include these ingredients and the EDGES measurement has inspired many new models. These explain the large amplitude using either an excess cooling of ordinary baryonic matter \cite{Bar_nat, Munoz, Berlin, Liu} or an excess radiowave background (ERB) \cite{Feng_2018, Fialkov_19}.

The motivation for the latter idea is the recently-confirmed excess radio brightness above the CMB by the Long Wavelength Array (LWA1) \cite{Dowell_2018} for frequencies 40--$\SI{80}{\mega\hertz}$. This excess was first detected by Absolute Radiometer for Cosmology, Astrophysics and Diffuse Emission (ARCADE 2) \cite{Fixsen_2011}. While the origin of this excess radio background is not very well understood \cite{Singal_2018}, it still provides an empirically plausible means of explaining the EDGES result.

Beyond these modelling efforts that aim to understand the EDGES result are the attempts to use the EDGES result for constraining various physical processes. Thanks to the richness of 21-cm physics, several studies on 21-cm signal have been theorised (many of which have existed even before the EDGES detection) to be useful to probe primordial magnetic fields \cite{Tashiro06, Dom09, Kunze_2019, Kunze19, Bera, natwar}, gravitational waves \cite{Somnath, Hirata18, Mishra}, warm dark matter \cite{Safarzadeh_2018, wdm, atrideb2, boyarsky, vipp}, dark matter decay and annihilation \cite{PhysRevD.74.103502, valdes, damico, Liu18}, viscous dark matter \cite{bhatt} and many other exotic processes. We address one such possibility in this paper by considering the effect of primordial black holes (PBHs) on the 21-cm signal and deriving constraints on the abundance of PBHs from the EDGES measurement.

Understanding the constituents of dark matter (DM) is one of the major problems in cosmology. Primordial black holes, possibly formed due to the collapse of large density perturbations in the very early Universe~\cite{1966AZh....43..758Z,Hawking:1971ei,chapline, Khlopov_2010}, can partly or even entirely explain the present day DM density~\cite{carr, Green_2021, bernard, villa, dasgupta}. Ultralight evaporating PBHs, i.e., PBHs in the mass range of $\num{e15}$--$\SI{e17}{\gram}$ are typically probed via observations of their evaporation products. Non-observations of such Hawking radiated photons~\cite{BJC,Arbey:2019vqx,Ballesteros:2019exr,integral,comptel, iguaz2021}, neutrinos~\cite{dasgupta}, and electrons/positrons~\cite{voyager,511kev2,511kev1,dasgupta, Cai_2021} in various space- as well as ground-based detectors provide stringent exclusion limits on the fraction of DM composed of ultralight evaporating PBHs. PBHs in these mass range can also be constrained via precise measurements of the CMB~\cite{CMBevap,Stocker:2018avm,Poulter:2019ooo,Acharya:2020jbv, Cang_2021}, dwarf galaxy heating~\cite{Kim:2020ngi,Laha:2020vhg}, and radio observations~\cite{Chan:2020zry,Dutta:2020lqc}. Observations of low energy ($\sim\si{\mega\electronvolt}$) photons from the Galactic Centre by the upcoming soft gamma-ray telescopes such as AMEGO, as well as next-generation neutrino detectors can also project stringent exclusion limits on the fraction of DM composed of ultralight evaporating PBHs \cite{Ray:2021mxu,Wang:2020uvi,Calabrese:2021zfq,DeRomeri:2021xgy}. The measurement of the 21-cm signal is considered to be another promising probe that can be used to constrain the DM fraction of PBHs in various mass windows~\cite{mack, hektor, Clark, mena, Yang, Halder, Tashiro, Tash}. Recently, by using the EDGES measurement, Clark et al. (2018)~\cite{Clark} showed that evaporating PBHs in the mass range of $\num{e15}$--$\SI{e17}{\gram}$ cannot form the solitary component of DM.

In this paper, we revisit 21-cm constraints on the PBH abundance and aim to improve upon previous work \cite{Clark} in two significant ways. Firstly, we use a more sophisticated model of all the known astrophysical phenomena that affect the 21-cm signal. This includes processes such as X-ray heating, Ly~$\alpha$ heating, and an ERB that were neglected in previous work, but can nonetheless have a significant impact on the inferred PBH constraints. Secondly, and more importantly, we aim to utilise all of the information contained in the EDGES measurement in our analysis. Previously published analysis \cite{Clark} only used a central location of the 21-cm absorption feature. In this work we use the full EDGES data for $28\gtrsim z\gtrsim14$, rather than just at a specific redshift, thus capturing all features of the observed signal. We use an MCMC-enabled Bayesian analysis to derive our constraints. This also allows us to study the covariance of PBH parameters with other processes affecting the 21-cm signal.

This paper is organized as follows. In section~\ref{theory}, we describe our model of the 21-cm signal. Section~\ref{pbh} discusses the effect of PBHs on the 21-cm signal. We describe our inference procedure in section~\ref{method}. Our results are presented in section~\ref{results}. We discuss the implications of our results and some caveats in section~\ref{disc}, and end with a summary in section~\ref{Conc}. The following cosmological parameters are used: $\Omega_{\mathrm{m}}= 0.315$, $\Omega_{\mathrm{b}}=0.049$, $\Omega_\Lambda = 0.685$, $h=0.674$, $Y_{\mathrm{p}}=0.245$, $T_0=\SI{2.725}{\kelvin}$, $\sigma_8 = 0.811$ and $n_{\mathrm{s}} = 0.965$ \cite{Fixsen_2009, Planck}, where $T_0$ and $Y_{\mathrm{p}}$ are the CMB temperature measured today and primordial helium fraction by mass, respectively.

\section{The global 21-cm signal}\label{theory}

The global (sky-averaged) 21-cm signal is the 21-cm brightness measured against the background (CMB or CMB+ERB, as in this work). Because of the long wavelength, the intensities can be written in terms of temperatures using the Rayleigh-Jeans law giving rise to a `differential brightness temperature' for the global 21-cm signal \cite{Furlanetto, Pritchard_2012, BARKANA_2016, Mesinger_19,Liu_2020}, given by
\begin{equation}
\Delta T_\mathrm{b}=27\bar{x}_{\textsc{Hi}}\left(\frac{1-Y_{\mathrm{p}}}{0.76}\right)\left(\frac{\Omega_\mathrm{b}h^2}{0.023}\right)\sqrt{\frac{0.15}{\Omega_\mathrm{m}h^2}\frac{1+z}{10}}\left(1-\frac{T_{\mathrm{r}}}{T_\mathrm{s}}\right)\si{\milli\kelvin}\,,\label{DeltaT}
\end{equation} 
where $x_{\textsc{Hi}}$ is the neutral hydrogen fraction, $T_\mathrm{s}$ is the spin temperature, and $T_{\mathrm{r}}$ is the net background temperature.

The spin temperature is not a real thermodynamic quantity but only an effective excitation temperature that quantifies the relative population of the hyperfine levels \cite{Ewen}. The processes affecting these populations and hence the spin temperature are the three Einstein processes, the Wouthuysen-Field effect \cite{Wouth, Field}, and the collision of hydrogen atoms with free electrons and other hydrogen atoms. Detailed balance between these processes then gives us the spin temperature
\begin{equation}
T_\mathrm{s}^{-1}=\frac{x_{\mathrm{r}} T_{\mathrm{r}}^{-1}+(x_\mathrm{k}+x_\alpha)T_\mathrm{k}^{-1}}{x_{\mathrm{r}}+x_\mathrm{k}+x_\alpha}\,,\label{spin_temp}
\end{equation}
where $x_{\mathrm{r}}$, $x_\mathrm{k}$ and $x_\alpha$ are the 21-cm, collisional and Ly~$\alpha$ coupling, respectively. Because of a near thermal equilibrium of gas and Ly~$\alpha$ photons, we have made an assumption that the colour temperature is equal to the gas kinetic temperature, i.e. $T_\alpha\approx T_{\mathrm{k}}$ \cite{Field}. The modelling of hyperfine line couplings, $x_{\mathrm{r}}$, $x_\mathrm{k}$ and $x_\alpha$, is discussed in section~\ref{hlc}, and that for the evolution of the gas kinetic temperature $T_\mathrm{k}$ is given in section~\ref{gastemp}, where we outline our model for X-ray heating and Ly~$\alpha$ heating that were previously ignored.

In eq.~\eqref{DeltaT}, $x_{\textsc{Hi}}$ is the neutral hydrogen fraction. The bar represents a global average over the cosmic volume, which includes \textsc{H\,i} and \textsc{H\,ii} regions. However, in this work we will work at redshifts where the effects of reionization are unimportant, in which case the ionized volume fraction of \textsc{H\,ii} region is always zero, i.e., $Q_{\textsc{Hii}}=0$. This simplifies the calculation of $\bar{x}_{\textsc{Hi}}$ and we write~\cite{Mesinger_11, Mesinger_13}
\begin{equation}
\bar{x}_{\textsc{Hi}}=(1-Q_{\textsc{Hii}})(1-x_\mathrm{e})=1-x_{\mathrm{e}}\,,\label{xHI}
\end{equation}
where $x_\mathrm{e}$ is the electron fraction defined as the number density of electrons relative to total hydrogen
\begin{equation}
x_{\mathrm{e}}\equiv \frac{n_{\mathrm{e}}}{n_{\mathrm{H}}}\,.
\end{equation}
in the IGM. We discuss our model for $x_\mathrm{e}$ in section~\ref{ion}.

The background temperature $T_{\mathrm{r}}$ includes the standard contribution from the CMB as well as a possible ERB, as we discuss in section~\ref{ERB}.

\subsection{Hyperfine line couplings}\label{hlc}

We now discuss our models for the hyperfine line couplings, $x_{\mathrm{r}}$, $x_\mathrm{k}$ and $x_\alpha$.

\subsubsection{21-cm coupling}
The 21-cm coupling is given by \cite{Venu}
\begin{equation}
x_{\mathrm{r}}=\frac{1-\ue^{-\tau_{21\mathrm{cm}}}}{\tau_{21\mathrm{cm}}}\,,\label{xgamma}
\end{equation}
where
\begin{equation}
\tau_{21\mathrm{cm}}=\frac{3}{32\pi}\frac{A_{10}}{H}n_{\textsc{Hi}}\lambda_{21\mathrm{cm}}^3\frac{T_*}{T_\mathrm{s}}\,,\label{tau21}
\end{equation}
is the 21-cm optical depth \citep[e.g.,][]{BL05}. Because $T_\mathrm{s}$ and $x_{\mathrm{r}}$ are dependent on each other, we may find their values iteratively as follows \cite{Fialkov_19}:
\begin{enumerate}
\item[1.] Set $x_{\mathrm{r}}=1$
\item[2.] Evaluate $T_\mathrm{s}$ using eq.~\eqref{spin_temp}
\item[3.] Using $T_\mathrm{s}$ evaluated in step 2, find the new $x_{\mathrm{r}}$ using eqs.~\eqref{xgamma} and \eqref{tau21}
\item[4.] Repeat from step 2 using an updated $x_{\mathrm{r}}$ found in step 3
\end{enumerate}
The convergence is rapid and usually 3 iterations are sufficient. The resultant value of $x_{\mathrm{r}}$ is usually close to 1.

\subsubsection{Collisional coupling}
Collisions can cause hyperfine transition in a neutral hydrogen atom via two different mechanisms: (\textit{a}\/) by spin exchange in which collisions with other hydrogen atoms, electrons, or protons swap the electron with another that has the opposite spin, or (\textit{b}\/) by spin flip of the electron via magnetic forces. Process (\textit{a}\/) is more likely to occur \cite{Furlanetto}. The collisional coupling is
\begin{equation}
x_{\mathrm{k}}=\frac{T_*C_{10}}{T_{\mathrm{r}} A_{10}}\,,\label{xk}
\end{equation}
where $C_{10}$ is the de-excitation rate by collisions and $A_{10}=\SI{2.85e-15}{\hertz}$ is the Einstein coefficient of spontaneous emission for the hyperfine transition \cite{Wild} and $T_*=h_{\mathrm{P}}\nu_{\mathrm{21cm}}/k_{\mathrm{B}}$ for Planck's constant $h_{\mathrm{P}}$ and Boltzmann constant $k_{\mathrm{B}}$. Collisional de-excitation rate is expressed as
\begin{equation}
C_{10}=n_{\textsc{Hi}}\kappa_{\mathrm{HH}}+n_{\mathrm{e}}\kappa_{\mathrm{eH}}+n_{\mathrm{p}}\kappa_{\mathrm{pH}}\,,
\end{equation}
where $n_{i}$ is the number density of species $i$ and $\kappa_{i\mathrm{H}}$ is the reaction rate, in units of volume per unit time, between $i$ and \textsc{H\,i}. Several papers \cite{Allison, Zygelman_2005,Furlanetto, FF06} have tabulated these rates at different temperatures. Useful fitting functions exist in literature which fit the data given in these tables. They are as follows \cite{Liszt, Kuhlen}:
\begin{equation}
\log_{10}\kappa_{\mathrm{eH}}=
\begin{cases}
-15.607+\frac{1}{2}\log_{10} T_{\mathrm{k}}\cdot\exp\left[{-(\log_{10} T_\mathrm{k})^{4.5}}/{1800}\right] \text{ if }T_\mathrm{k}<\num{e4}\\
-14.102\text{ if }T_\mathrm{k}\geqslant\num{e4}\,,
\end{cases}
\end{equation}
and
\begin{equation}
\kappa_{\mathrm{HH}}=\num{3.1e-17}T_{\mathrm{k}}^{0.357}\ue^{-32/T_{\mathrm{k}}}\,.    
\end{equation}
No fitting function for $\kappa_{\mathrm{pH}}$ is available in the literature.  We therefore construct and use the fit
\begin{equation}
\kappa_{\mathrm{pH}}=\num{e-16}\left[c_0+c_1\log_{10} T_\mathrm{k}+c_2\log_{10}^2 T_\mathrm{k}+c_3\log_{10}^3 T_\mathrm{k}\right]\,,
\end{equation}
where $c_0=4.28, c_1=0.24, c_2=-1.37$ and $c_3=0.53$ for the available data \cite{FF07}. All $\kappa_{i\mathrm{H}}$s are in $\si{\metre^3\second^{-1}}$. The final expression of $x_\mathrm{k}$ can be written as
\begin{equation}
x_{\mathrm{k}}=\frac{T_*n_\mathrm{H}}{T_\mathrm{r} A_{10}}\left[(1-x_{\mathrm{e}})\kappa_{\mathrm{HH}}+x_{\mathrm{e}}\kappa_{\mathrm{eH}}+x_{\mathrm{e}}\kappa_{\mathrm{pH}}\right]\,,\label{XK}
\end{equation}
where we have used the charge neutrality of the Universe by which $n_{\mathrm{e}}=n_{\mathrm{p}}$. 

\subsubsection{\texorpdfstring{Ly~$\alpha$ coupling}{Ly~α coupling}}\label{lyacoupling}
The Ly~$\alpha$ photons produced by the first galaxies indirectly affect the spin temperature through a process known as the Wouthuysen-Field effect \cite{Wouth, Field}. Accurate modelling of this coupling is essential at Cosmic Dawn since it is the Ly~$\alpha$ coupling that makes the 21-cm signal observable. The expression for the Ly~$\alpha$ coupling can be written as
\begin{equation}
x_\alpha=(1-\delta_\alpha)\frac{J_\alpha}{J_0}\,,\label{xalpha}
\end{equation}
where $\delta_\alpha$ represents a distortion in the Ly~$\alpha$ background due to its interaction with neutral hydrogen atom \cite{Mittal}
\begin{equation}
\delta_\alpha={}_{\phantom{1}3}F_0(1/3,2/3,1;0;-\xi_1)\,,\label{S}
\end{equation}
for
\begin{equation}
\xi_1=\frac{9\pi}{4a\tau_\alpha\eta^3}\,,
\end{equation}
and${}_{\phantom{1}3}F_0$ being the $(3,0)$-hypergeometric function \cite{ARFKEN}. The Voigt parameter \cite{Rybicki1994}, the Ly~$\alpha$ optical depth \cite{GP} and the recoil parameter \cite{Chen} are given by
\begin{subequations}\label{atn}
\begin{align}
a&=\frac{A_\alpha}{4\pi\nu_\alpha}\sqrt{\frac{m_{\mathrm{H}}c^2}{2k_\mathrm{B}T_\mathrm{k}}}\,,\\
\tau_{\alpha}&=\frac{3}{8\pi}\frac{A_\alpha}{H}n_{\textsc{Hi}}\lambda_{\alpha}^3\,,\\
\eta&=\frac{h_{\mathrm{P}}/\lambda_\alpha}{\sqrt{2m_{\mathrm{H}}k_\mathrm{B}T_\mathrm{k}}}\,,\label{eta}
\end{align}
\end{subequations}
respectively. Here $A_\alpha=\SI{6.25e8}{\hertz}$ is the Einstein spontaneous emission coefficient of Ly~$\alpha$ transition, $m_\mathrm{H}$ is the mass of hydrogen, $\lambda_\alpha(\nu_\alpha)$ is the wavelength (frequency) of the Ly~$\alpha$ photon, and $c$ is the speed of light.

The factor $J_0$ is a combination of fundamental constants and background temperature \cite{Mittal}
\begin{equation}
J_{0}=\num{5.54e-8}\frac{T_{\mathrm{r}}}{T_0}\,\si{\per\metre\squared\per\second\per\hertz\per\steradian}\,.
\end{equation}

To calculate the undisturbed Ly~$\alpha$ specific intensity far from the resonance line, $J_\alpha$, we need the comoving emissivity. The latter is defined as the number of photons emitted per unit comoving volume per unit proper time per unit energy at redshift $z$ and energy $E$. It can be constructed based on the approach taken in ref.~\cite{BL05}. We choose Population II type star as our base model for spectral energy distribution (SED, number of photons emitted per unit energy per baryon). It is given by \cite{Mittal}
\begin{equation}
\phi_{\alpha}(E)=
\begin{cases}
2902.91\, \hat{E}^{-0.86}&\text{if }E\in[E_\alpha,E_\beta]\\
1303.34\, \hat{E}^{-7.66}&\text{if }E\in(E_\beta,E_\infty]\,,
\end{cases}
\end{equation}
in $\si{\per\electronvolt}$, where $\hat{E}=E/E_\infty$, $E_\alpha=\SI{10.2}{\electronvolt}$, $E_\beta=\SI{12.09}{\electronvolt}$ and $E_\infty=\SI{13.6}{\electronvolt}$ are the energies corresponding to Ly~$\alpha$, Ly~$\beta$ and Lyman limit transition, respectively.

We can now write the comoving emissivity as
\begin{equation}
\epsilon_\alpha(E,z)=f_\alpha\phi_{\alpha}(E)\frac{\dot{\rho}_\star(z)}{m_{\mathrm{b}}}\,,\label{epsilon}
\end{equation}
where $\dot\rho_\star$ is the comoving star formation rate density (SFRD) and $m_{\mathrm{b}}$ is the number-averaged baryon mass given as \cite{Haimoud}
\begin{equation}
m_{\mathrm{b}}=\frac{m_{\mathrm{H}}n_{\mathrm{H}}+m_{\mathrm{He}}n_{\mathrm{He}}+m_\mathrm{e}n_{\mathrm{e}}}{n_{\mathrm{H}}+n_{\mathrm{He}}+n_{\mathrm{e}}}\,.
\end{equation}
Neglecting the mass of electron and using
\begin{equation}
x_\mathrm{He}\equiv \frac{n_{\mathrm{He}}}{n_{\mathrm{H}}}=\frac{Y_{\mathrm{p}}}{4(1-Y_{\mathrm{p}})}\,,\label{xHe}
\end{equation}
which is number density of helium relative to hydrogen, we get
\begin{equation}
m_{\mathrm{b}}=\frac{4m_{\mathrm{H}}}{4-3Y_{\mathrm{p}}+4x_{\mathrm{e}}(1-Y_{\mathrm{p}})}\,.\label{abm}
\end{equation}
Since $x_{\mathrm{e}}$ is quite small, typically $\sim\mathcal{O}(\num{e-3})$ for the redshift range considered in this work, we can write $m_{\mathrm{b}}\approx1.22m_{\mathrm{H}}$.

The comoving SFRD, represented by $\dot{\rho}_\star(z)$, and measured in mass per unit time per unit comoving volume, is set by the rate at which baryons collapse into dark matter haloes \cite{BL05}. We will assume that only haloes of virial temperatures $(T_{\mathrm{vir}})$ above a certain given value will contribute. Their number at a given redshift can be determined by the Press-Schechter formalism \cite{Press}. Thus,
\begin{equation}
\dot{\rho}_\star(z)=-f_\star\bar{\rho}_\mathrm{b}^0 (1+z)H(z)\frac{\ud F_{\mathrm{coll}}(z)}{\ud z}\,,\label{fstar}
\end{equation}
where 
\begin{equation}
\bar{\rho}_\mathrm{b}^0=\frac{3H_0^2}{8\pi G_{\mathrm{N}}}\Omega_\mathrm{b}\,,
\end{equation}
is the mean cosmic baryon mass density measured today ($H_0$ is the Hubble's constant measured today and $G_{\mathrm{N}}$ is the Newton's gravitational constant), $f_\star$ is the star formation efficiency, defined as the fraction of baryons converted into stars in the haloes. Because it is completely degenerate with $f_{\alpha}$ and $f_{\mathrm{X}}$ (to be introduced in section~\ref{fx}) it does not matter what value we choose for it. Here we take the value from the fiducial set of parameters for Pop II stars from ref.~\cite{F06}, i.e. $f_\star=0.1$. We denote the fraction of baryons that have collapsed into dark matter haloes by $F_{\mathrm{coll}}$, given by \cite{BL01}
\begin{equation}
F_{\mathrm{coll}}(z)=\mathrm{erfc}\left[\frac{\delta_{\mathrm{crit}}(z)}{\sqrt{2}\sigma(m_{\mathrm{min}})}\right]\,,
\end{equation}
where erfc($\cdot$) represents the complementary error function, $\delta_{\mathrm{crit}}$ is the linear critical density of collapse and $\sigma^2$ is the variance in smoothed density field. The minimum virial temperature enters the model through the expression of minimum halo mass for star formation, i.e.,
\begin{equation}
m_{\mathrm{min}}=10^8\frac{1}{\sqrt{\Omega_{\mathrm{m}}h^2}}\mathrm{M}_{\odot}\left[\frac{10}{1+z}\frac{0.6}{\mu}\frac{\mathrm{min}(T_{\mathrm{vir}})}{\num{1.98e4}}\right]^{3/2}\,,\label{tvir}
\end{equation}
where $h$ is the Hubble's constant measured today in units of $\SI{100}{\kilo\metre\per\second\per\mega\parsec}$ and $\mu\approx1.22$ \cite{DAYAL20181}. We calculate $\delta_{\mathrm{crit}}(z)/\sigma(m_{\mathrm{min}})$ using \texttt{COLOSSUS}\footnote{\url{https://bitbucket.org/bdiemer/colossus/src/master/}} \cite{Colossus}. As an example, for our cosmological parameters and $\mathrm{min}(T_{\mathrm{vir}})=\SI{e4}{\kelvin}$ we get $F_{\mathrm{coll}}(z=0)\approx0.735$.

We treat the minimum $\log_{10}T_{\mathrm{vir,}4}$, which is a shorthand for $\log_{10} [T_{\mathrm{vir}}/(\SI{e4}{\kelvin})]$, as a free parameter, and vary it between $-0.75$ and $1.25$. For base model we take it to be 0, which corresponds to the atomic cooling threshold \cite{BL01}. Our choice of the range is justified as it covers a large range of possible 21-cm signal values. See figure~\ref{comb} for effect of changing this parameter on gas kinetic temperature and the 21-cm signal.

We scale the Ly~$\alpha$ background up and down using the parameter $f_\alpha$. Since we are uncertain about the SED or the SFRD of the galaxies, we set up a basic conservative model and vary it using $f_\alpha$. In this work we vary it between 0.01 and 100 (cf. ref.~\cite{Mittal}) with the base value at 1. Figure~\ref{comb} shows the effect of varying this parameter on the 21-cm signal.

We can now evaluate $J_\alpha$ from $\epsilon_\alpha$ (after internally converting $\epsilon_\alpha$ from per unit energy basis to per unit frequency basis) as
\begin{equation}
J_\alpha(z)=\frac{c}{4\pi}(1+z)^2\sum_{n=2}^{23}P_n\int_z^{z_{\mathrm{max}}}\frac{\epsilon_{\alpha}(E_n',z')}{H(z')}\,\ud z'\,,\label{J}
\end{equation}
in units of number per unit time per unit area per unit frequency per unit solid angle. The $n^{\mathrm{th}}$ term in the sum accounts for the finite probability $P_n$ with which a photon in the upper Lyman line will redshift to Ly~$\alpha$ wavelength. The values of $P_n$ are computed in an iterative fashion using the selection rule and the decay rates. The detailed procedure and table of values can be found in refs.~\cite{Hirata, PF06}. The redshifted energy of $n^{\mathrm{th}}$ Lyman series line is given by
\begin{equation}
E_n'=E_n\frac{1+z'}{1+z}\,, 
\end{equation}
where $E_n$ is the energy of the photon released in transition from $n^{\mathrm{th}}$ state to ground state
\begin{equation}
E_n=13.6\left(1-\frac{1}{n^2}\right)\,\si{\electronvolt}\,.
\end{equation}
The maximum redshift from which this photon could have been received is given by
\begin{equation}
1+z_{\mathrm{max}}=\frac{E_{n+1}}{E_{n}}(1+z)=\frac{1-(1+n)^{-2}}{1-n^{-2}}(1+z)\,.    
\end{equation}

\subsection{Gas kinetic temperature and heating processes}\label{gastemp}

The evolution of the gas kinetic temperature is also important in setting the amplitude of the 21-cm signal through the spin temperature (eq.~\ref{spin_temp}). This is given by \cite{Mittal}
\begin{equation}
(1+z)\frac{\ud T_{\mathrm{k}}}{\ud z}=2T_{\mathrm{k}}-\frac{T_{\mathrm{k}}(1+z)}{1+x_\mathrm{He}+x_\mathrm{e}}\frac{\ud x_\mathrm{e}}{\ud z}-\frac{2}{3n_{\mathrm{b}}k_{\mathrm{B}}H}\sum q\,,\label{DEofTk}
\end{equation}
where $n_{\mathrm{b}}=n_{\mathrm{H}}(1+x_{\mathrm{He}}+x_{\mathrm{e}})$ is the total particle number density. The first term on the right hand side of eq.~\eqref{DEofTk} is the adiabatic cooling term because of an adiabatically expanding Universe, second term accounts for the change in internal energy due to changing particle number, and finally the third term is the sum of all heating and cooling processes. In this work we consider Compton, Ly~$\alpha$, Hawking radiation (HR), and X-ray heating. They are discussed one by one next. We must couple eq.~\eqref{DEofTk} with an equation for the variation of $x_{\mathrm{e}}$. The latter is discussed in section~\ref{ion}. The redshifts of our interest range from $1+z=60$ to $1+z=14$ with the initial condition, obtainable from \texttt{RECFAST},\footnote{\url{https://www.astro.ubc.ca/people/scott/recfast.html}.} being $x_{\mathrm{e}}\approx\num{2.47e-4}$ and $T_{\mathrm{k}}\approx\SI{70.28}{\kelvin}$ at $1+z=60$.

\subsubsection{Compton heating}
After recombination, the inverse Compton scattering of electrons off the background photons couples the matter and background radiation for $z\gtrsim200$. Because of this Compton heating the temperature of matter and the background fall together as $(1+z)$ at these redshifts. As the Universe expands and becomes neutral Compton heating ceases to play any role \cite{Seager_1999, Seager_2000}. However, because of partial X-ray ionization of IGM in later times, $z\lesssim60$, it may have a finite contribution, although only a minor one. The Compton heating term is \cite{Weymann}
\begin{equation}
\frac{2q_{\mathrm{Comp}}}{3n_{\mathrm{b}}k_\mathrm{B}H}=\frac{32\sigma_\mathrm{T}\sigma_\mathrm{S} T_\gamma^4}{3Hm_{\mathrm{e}}c^2}\frac{1}{1+x_{\mathrm{He}}+x_{\mathrm{e}}}(T_\gamma-T_\mathrm{k})\,,
\end{equation}
where $T_\gamma=T_0(1+z)$ is the CMB temperature at a redshift $z$, $\sigma_\mathrm{T}=\SI{6.65e-29}{\metre^2}$ is the Thomson scattering cross section and $\sigma_{\mathrm{S}}=\SI{5.67e-8}{\watt\metre^{-2}\kelvin^{-4}}$ is the Stephan-Boltzmann constant.

\subsubsection{\texorpdfstring{Ly~$\alpha$ heating}{Ly~α heating}}
The other heating process we will consider here is the Ly~$\alpha$ heating \cite{Mittal}, which is a `side effect' of Ly~$\alpha$ coupling. Following our previous work \cite{Mittal}, the term to be added in eq.~\eqref{DEofTk} is
\begin{equation}
\frac{2q_{\alpha}}{3n_{\mathrm{b}}k_{\mathrm{B}}H}=\frac{8\pi}{3}\frac{h_{\mathrm{P}}}{\lambda_\alpha^2}\sqrt{\frac{2T_\mathrm{k}}{m_\mathrm{H}k_\mathrm{B}}}\frac{J_\alpha}{n_{\mathrm{b}}} \left(I_\mathrm{c}+\frac{J_\alpha^\mathrm{i}}{J_\alpha^{\mathrm{c}}} I_\mathrm{i}\right)\,.\label{qalpha}
\end{equation}
where $I_\mathrm{c}$ is the integral over the spectrum of continuum photons given as \cite{FP06}
\begin{equation}
I_\mathrm{c}=\eta(2\pi^4a^2\tau_\alpha^2)^{1/3}\left[\mathrm{Ai}^2(-\xi_2)+\mathrm{Bi}^2(-\xi_2)\right]\,,
\end{equation}
where
\begin{equation}
\xi_2=\eta\left(\frac{4a\tau_\alpha}{\pi}\right)^{1/3}\,,
\end{equation}
Ai and Bi represent the Airy function of first and second kind, respectively. See eq.~\eqref{atn} for $a,\tau_{\alpha}$ and $\eta$. Similarly, $I_{\mathrm{i}}$ is the integral over the spectrum of injected photons given as \cite{Mittal}
\begin{equation}
I_\mathrm{i}=\eta\sqrt{\frac{a\tau_\alpha}{2}}\int_{0}^{\infty}\left[\exp\left(-2\eta y-\frac{\pi y^3}{6a\tau_\alpha}\right)\mathrm{erfc}\left(\sqrt{\frac{\pi y^3}{2a\tau_\alpha}}\right)\frac{1}{\sqrt{y}}\right]\,\ud y-\frac{\delta_\alpha(1-\delta_\alpha)}{2\eta}\,.\label{cool}
\end{equation}
with $\delta_\alpha$ given in eq.~\eqref{S}. The quantity $J_\alpha^\mathrm{i}/J_\alpha^{\mathrm{c}}$ is decided by the stellar model. Here we take it to be 0.2.

Note that the free parameter introduced in section~\ref{lyacoupling}, $f_{\alpha}$, directly affects $q_{\alpha}$ through $J_{\alpha}(z)$ which in turn is calculated from emissivity, eq.~\eqref{epsilon}. This parameter has a small influence on the thermal history but a more dramatic effect on the 21-cm signal. See figure~\ref{comb}.


\subsubsection{X-ray heating}\label{fx}
Ultraviolet (UV) and X-ray photons from high-redshift galaxies ionize and heat the IGM. Due to the large cross-section of hydrogen atoms at UV wavelengths, the UV photons have short mean free path and are mainly responsible for ionizing the medium in close vicinity of the sources resulting in the so-called \textsc{H\,ii} regions \cite{BL01, Wyithe_2003, BL04}. X-rays have very long mean free path, because of which they are able to penetrate far into the IGM. In the process, they heat and partially ionize the IGM \cite{Mirabel}. The possible sources of X-rays include X-ray binaries \cite{Power_2013, Fragos_2013}, inverse Compton scattering in supernova remnants \cite{Oh_2001} and mini-quasars \cite{Madau_2004}. X-ray binaries are a class of binary stars that are luminous in X-rays due to accretion from one of the stars onto another \cite{tauris}. Evidence suggests that at high redshifts, such as those considered in this work, the dominant source could be high-mass X-ray binaries \cite{Fabbiano, Brorby2016, Lehmer2016}. In canonical models of the 21-cm absorption feature at Cosmic Dawn, the Ly~$\alpha$ photons create the absorption feature and X-ray photons destroy it.  

X-ray heating is commonly characterised by three parameters, namely $w, E_0$ and $f_{\mathrm{X}}$. These represent the power law index of the X-ray background spectral energy distribution (SED), the minimum energy of X-ray photons that can contribute to heating, and an overall normalisation of X-ray SED, respectively \cite{Cohen17, Fialkov_19}. In this work we keep $w$ and $E_0$ fixed and vary only $f_{\mathrm{X}}$. This is justified because the dependence of 21-cm signal on $w$ is weak and $E_0$ is somewhat degenerate with $f_{\mathrm{X}}$ \cite{Monsalve_2019}.

The observed relationship between the X-ray luminosity $L_{\mathrm{X}}$ and the star formation rate (SFR) in star-forming galaxies suggests a linear relationship \cite{Grim03, Gilfanov}, which can be written as \cite{Mineo} 
\begin{equation}
\frac{L_{\mathrm{X}}}{\mathrm{SFR}}\approx\SI{2.61e32}{\watt}\left(\mathrm{M}_{\odot}\mathrm{yr}^{-1}\right)^{-1}\,,\label{LxSFR}
\end{equation}
for photon energies 0.5--$\SI{8}{\kilo\electronvolt}$ and a power law SED with $w\sim1.5$. We extrapolate this relation over 0.2--$\SI{30}{\kilo\electronvolt}$ assuming the same power law. The reason for extrapolating to lower energies is because of large cross-section of X-ray and neutral hydrogen interaction, which roughly goes as $\sigma(E)\propto E^{-3}$ but too low energy photons with $E<\SI{0.2}{\kilo\electronvolt}$ are excluded since they are absorbed into IGM over short distances from the source \cite{F06, Mirocha_14}. Higher energy photons with $E>\SI{30}{\kilo\electronvolt}$ have longer mean free path but smaller cross section, which means they have a negligible contribution in heating. Thus, 0.2--$\SI{30}{\kilo\electronvolt}$ seems to be a reasonable choice \cite{Mirocha_19}.

The SED (in units of number per unit energy per baryon) of X-rays is \cite{Mesinger_11, Mesinger_13},
\begin{equation}
\phi_{\mathrm{X}}(E)=\frac{N_{\mathrm{X}}}{E_0}\frac{w-1}{1-(E_0/E_1)^{w-1}}\left(\frac{E}{E_0}\right)^{-w-1}\,,\label{phix}
\end{equation}
where $w=1.5$ as already mentioned before, $N_{\mathrm{X}}$ is the number of X-ray photons emitted per stellar baryon, $E_0=\SI{0.2}{\kilo\electronvolt}$ and $E_1=\SI{30}{\kilo\electronvolt}$ are the minimum and maximum X-ray energy relevant for heating. We get $N_{\mathrm{X}}\sim 1$ on extrapolating 0.5--$\SI{8}{\kilo\electronvolt}$ $L_{\mathrm{X}}$-SFR relation (given in eq.~\ref{LxSFR}) to 0.2--$\SI{30}{\kilo\electronvolt}$.

In analogy with Ly~$\alpha$ emissivity, we can now construct X-ray emissivity as follows (in units of number per unit time per unit energy per unit comoving volume) \cite{Mesinger_11, Mesinger_13}
\begin{equation}
\epsilon_{\mathrm{X}}(E,z)=f_{\mathrm{X}}\phi_{\mathrm{X}}(E)\frac{\dot{\rho}_\star(z)}{m_{\mathrm{b}}}\,,\label{epsilonx}
\end{equation}
where $m_{\mathrm{b}}$ is the average baryon mass (eq.~\ref{abm}) and $\dot{\rho}_\star$ is the comoving SFRD (eq.~\ref{fstar}). We will vary $f_{\mathrm{X}}$ between 0.1 and 10, with 1 as its base model value. This choice is consistent with previous studies \cite{Fialkov_17, Cohen17}. In figure~\ref{comb} we show how our gas temperature and 21-cm signal change when $f_{\mathrm{X}}$ is changed.

The microscopic mechanism of X-ray heating can be explained as follows \cite{F06, Madau_2017}. First, the X-rays photoionize the \textsc{H\,i} and He\,\textsc{i}. In this process hot energetic electrons are produced which dissipate their energy via atomic excitations, secondary ionizations, or collisions with other electrons. As a result the average kinetic energy, and hence the temperature of IGM, increases.

There are mainly two types of estimates for X-ray heating in literature. Some studies give this simply as a certain fraction of emissivity \cite{F06,Mirocha_2013, Mirocha_2015}. The other type is a more physically motivated version where it is calculated from the background specific intensity of X-rays, $J_{\mathrm{X}}$ \cite{Mesinger_11, Mesinger_13, Mirocha_14}. We use the second version with the detailed mathematical structure as follows. The standard photoheating rate is \cite{Madau_2017}
\begin{equation}
H_{\mathrm{X}}=4\pi\int_{E_0}^{E_1} (E-E_{\infty})\sigma(E)J_{\mathrm{X}}(E,z)\,\ud E\,,\label{Hx}
\end{equation}
where $E_{\infty}=\SI{13.6}{\electronvolt}$ is the ionization energy of hydrogen, and $\sigma(E)$ is the photoionization cross-section of \textsc{H\,i}--X-ray interaction, which takes the following functional form \cite{Verner}
\begin{equation}
\sigma(E)=\num{5.48e-18}\frac{(\mathcal{E}-1)^2}{(1+\sqrt{\mathcal{E}/32.88})^{2.96}}\mathcal{E}^{-4.02}\, \si{\metre^2}\,,\label{sigmae}
\end{equation}
for $\mathcal{E}=E/0.4298$ when $E$ is expressed in $\si{\electronvolt}$. Because of the energy division explained in previous paragraph, the standard photoheating rates are reduced. The reduced rate can be written as $f_{\mathrm{X,h}}H_{\mathrm{X}}$, where the reduction factor is given by \cite{Shull}
\begin{equation}
f_{\mathrm{X,h}}=1-\left(1-x_{\mathrm{e}}^{0.2663}\right)^{1.3163}\,.
\end{equation}
The final term to be inserted in eq.~\eqref{DEofTk} is
\begin{equation}
\frac{2q_{\mathrm{X}}}{3n_{\mathrm{b}}k_{\mathrm{B}}H}=\frac{8\pi}{3} \frac{1-x_{\mathrm{e}}}{1+x_{\mathrm{He}}+x_{\mathrm{e}}}\frac{f_{\mathrm{X,h}}}{k_{\mathrm{B}}H}\int_{E_0}^{E_1} (E-E_{\infty})\sigma(E)J_{\mathrm{X}}(E,z)\,\ud E\,.\label{qx}
\end{equation}

The background specific intensity of X-rays is $J_{\mathrm{X}}$, analogous to $J_{\alpha}$. We define it in terms of number per unit time per unit energy per unit area per unit solid angle. It can be calculated from the comoving X-ray emissivity of the source $\epsilon_{\mathrm{X}}$ (eq.~\ref{epsilonx}) as
\begin{equation}
J_{\mathrm{X}}(E,z)=\frac{c(1+z)^2}{4\pi}\int_z^{z_{\star}}\frac{\epsilon_{\mathrm{X}}(E',z')}{H(z')}e^{-\tau_{\mathrm{X}}(E,z,z')}\ud z'\,,\label{Jx}
\end{equation}
where $z_\star\sim60$ is the redshift when the star formation starts \cite{Mirocha_19}, and $E'=E(1+z')/(1+z)\,.$

The X-ray optical depth can be written as
\begin{equation}
\tau_{\mathrm{X}}(E,z,z')=\int_z^{z'}\frac{c\,\ud z''}{(1+z'')H(z'')\lambda_{\mathrm{X}}(E'',z'')}\,,
\end{equation}
where $E''=E(1+z'')/(1+z)$ and X-ray mean free path is approximately \cite{Furlanetto}
\begin{equation}
\lambda_{\mathrm{X}}(E,z)=\frac{1.1\bar{x}_{\textsc{Hi}}^{-1/3}}{(1+z)^3}\left(\frac{E}{\SI{300}{\electronvolt}}\right)^{3}\,\si{\giga\parsec}\,,\label{lambdax}
\end{equation}
in proper units. A better version of $\lambda_{\mathrm{X}}$ would be written in terms of a sum over number density and photoionization cross section for all species involved, such as the one in ref.~\cite{Mesinger_11}. But here we will continue to use the approximation in eq.~\eqref{lambdax}.

Note that the atomic excitations by the electrons released in photoionization can also generate Ly~$\alpha$ photons. This is modelled simply by saying that this extra Ly~$\alpha$ emissivity is a fraction of the X-ray emissivity. In this work, however, we have neglected that contribution (see for e.g. \cite{Reis2021}).

\subsection{Electron fraction and ionization rates}\label{ion}
The photons and electrons from Hawking emission also cause ionization of IGM by a mechanism similar to the X-ray photons and secondary ionization by electrons. The differential equation relevant for our work governing the evolution of electron fraction in the IGM is \cite{Madau_2017, Clark}
\begin{equation}
(1+z)H\frac{\ud x_{\mathrm{e}}}{\ud z}=\alpha(T_{\mathrm{k}})n_{\mathrm{H}}x^2_{\mathrm{e}}-\Gamma_{\mathrm{X}}(1-x_{\mathrm{e}})-\Gamma_{\mathrm{HR}}\,,\label{xe_bulk}
\end{equation}
where the first term on the right hand side is recombination term and the two negative terms are the ionization terms due to the X-ray background and Hawking radiation from PBHs. An approximate temperature dependence of recombination coefficient can be expressed as \cite{Theuns}
\begin{equation}
\alpha(T)=\num{2.5e-16}\frac{T^{-0.7}}{1+\left[T/(\SI{e6}{\kelvin})\right]^{0.7}}\,\si{\metre^3\second^{-1}}\,,
\end{equation}
with $T$ being in $\si{\kelvin}$.

The standard photoionization rate due to X-rays is given by
\begin{equation}
\Gamma_{\mathrm{X}}(z)=4\pi \int_{E_0}^{E_1} \sigma(E)J_{\mathrm{X}}(E,z)\,\ud E\,,\label{PIR}
\end{equation}
but it gets slightly enhanced due to secondary ionizations \cite{Madau_2017}. Thus, we make the following replacement
\begin{equation}
\Gamma_{\mathrm{X}}\to \Gamma_{\mathrm{X}}+\frac{f_{\mathrm{X,ion}}}{E_\infty}H_{\mathrm{X}}\,,
\end{equation}
where $H_{\mathrm{X}}$ was defined in eq.~\eqref{Hx}. The enhancement factor is \cite{Shull}
\begin{equation}
f_{\mathrm{X,ion}}=0.3908\left(1-x_{\mathrm{e}}^{0.4092}\right)^{1.7592}\,.
\end{equation}
For higher accuracy the factors $f_{\mathrm{X,h}}$ and $f_{\mathrm{X,ion}}$ may be used from ref.~\cite{F10}, but we will continue with the ones from ref.~\cite{Shull}.

We defer a discussion on ionization due to HR, i.e., $\Gamma_{\mathrm{HR}}$, in section~\ref{pbh}.

\subsection{Excess radio background}\label{ERB}
We have so far considered only the standard astrophysics. However, as mentioned in section~\ref{intro}, we need extra physical input in order to match the EDGES signal. We assume that there exists a uniform excess radio background in the sky. For radio frequencies the usual approximation of Rayleigh-Jeans limit works very well which allows us to quantify the energy flux in terms of a `temperature'. This radio flux is fit very well by a power law as observed by ARCADE~2 and LWA1 \cite{Fixsen_2011, Dowell_2018}. The combined CMB temperature and excess radio measured today at frequency $\nu$ can be written as
\begin{equation}
T_{\mathrm{r}}(\nu)=T_0+T_{\mathrm{R}}\left(\frac{\nu}{\nu_0}\right)^\beta\,,\label{excess0}
\end{equation}
where $T_0=\SI{2.725}{\kelvin}$ is the CMB temperature measured today,  $T_{\mathrm{R}}=\SI{24.1}{\kelvin}$, $\beta\approx-2.6$ \cite{Fixsen_2011} and reference frequency $\nu_0=\SI{310}{\mega\hertz}$. Generalising the above for an earlier epoch at redshift $z$ and measurement made for the frequency corresponding to 21-cm line (redshifted to $z$, i.e. $\nu=\nu_{\mathrm{21cm}}/(1+z)$) we get \cite{Feng_2018, Fialkov_19}
\begin{equation}
T_{\mathrm{r}}(z)=2.725(1+z)\left[1+0.169\,\zeta_{\mathrm{ERB}}(1+z)^{2.6}\right]\,,\label{excess}
\end{equation}
where we have parametrized the amplitude of excess radio --- the coefficient of the second term on the right hand side of eq.~\eqref{excess0} --- by $\zeta_{\mathrm{ERB}}$. We vary $\zeta_{\mathrm{ERB}}$ between 0.01 and 1 with $\zeta_{\mathrm{ERB}}=1$ as the base value corresponding to the excess observed by ARCADE 2. As with other parameters figure~\ref{comb} justifies the choice of this range.

Note that refs.~\cite{Ewall, Ewall2} considered an enhancement in the background, against which we measure our 21-cm brightness, due to the radiation emitted by accreting black holes of intermediate or supermassive black holes. However, ultralight PBHs in the mass range $\num{e15}$--$\SI{e17}{\gram}$ cannot produce photons of wavelength $\SI{21}{\centi\metre}$ either via accretion or via evaporation.\footnote{We find the brightness temperature at a photon energy of $\SI{5.9}{\micro\electronvolt}$ corresponding to the primary spectrum from a $\SI{e17}{\gram}$ PBH to be $\sim\SI{e-46}{\kelvin}$. However, we have not explored the same for secondary spectrum which may or may not contribute \cite{Mittal2021}.}

\section{Hawking radiation from primordial black holes}\label{pbh}
Primordial black holes emit particles via Hawking radiation and the spectrum of the emitted particles follows a blackbody like distribution. The emission rate of particles from a neutral and non-rotating PBH of temperature $T_{\mathrm{PBH}}$, in the energy interval $E$ and $E+\ud E$ is given by~\cite{Hawking, Page:1976df, Page:1976ki, gibbon, MacGibbon:1991tj, MacGibbon:2007yq}
\begin{equation}
\ud\dot{N}=\frac{1}{2\pi \hbar}\frac{\Gamma_s (E,\mu, T_{\mathrm{PBH}})}{\exp\left(\frac{E}{k_{\mathrm{B}}T_{\mathrm{PBH}}}\right)-(-1)^{2s}}\ud E\,,
\end{equation}
where $\hbar=h_{\mathrm{P}}/2\pi$ is the reduced Planck's constant, $s$ denotes the spin of the emitted particle and $\Gamma_s$ denotes a dimensionless absorption coefficient, commonly known as the greybody factor. For a neutral, and non-rotating PBH, the greybody factor depends on the energy of the emitted particle $E$, temperature of the PBH $T_{\mathrm{PBH}}$, spin of the emitted particle $s$, and the rest mass of the emitted particle $\mu$. In the high energy limit ($G_{\mathrm{N}}M_{\mathrm{PBH}}E/\hbar c^3 \gg 1$), the greybody factor becomes independent of the spin of the emitted particle and reaches its geometrical saturation value, i.e.\,$\Gamma= 27\,G^2_{\mathrm{N}} E^2(M^2_{\mathrm{PBH}} - \mu^2)/\hbar^2 c^6$. Whereas, in the opposite regime ($G_{\mathrm{N}}M_{\mathrm{PBH}}E/\hbar c^3 \ll 1$), it has a very strong dependence on the spin of the emitted particles~\cite{MacGibbon:2007yq,gibbon}. 

The temperature of an uncharged, non-rotating PBH is solely determined by its mass $M_{\mathrm{PBH}}$~\cite{Hawking, Page:1976df, Page:1976ki, gibbon, MacGibbon:1991tj, MacGibbon:2007yq} 
\begin{equation}
k_{\mathrm{B}}T_{\mathrm{PBH}}=\frac{\hbar c^3}{8\pi G_{\mathrm{N}} M_{\mathrm{PBH}}} = 1.06 \left(\frac{\SI{e13}{\gram}}{M_{\mathrm{PBH}}}\right) \si{\giga\electronvolt}\,.
\end{equation}
As the energy of the emitted particles becomes comparable to the temperature of a PBH, i.e.\,$E \sim k_{\mathrm{B}}T_{\mathrm{PBH}}$, significant Hawking emission occurs. Quantitatively, Hawking emission peaks at $E_{\mathrm{peak}}= 2.81\,k_{\mathrm{B}} T_{\mathrm{PBH}}$ for $s=0$ particle species, $E_{\mathrm{peak}}= 4.02 \,k_{\mathrm{B}}T_{\mathrm{PBH}}$ for $s=1/2$ particle species, and $E_{\mathrm{peak}}= 5.77\, k_{\mathrm{B}}T_{\mathrm{PBH}}$ for $s=1$ particle species~\cite{MacGibbon:2007yq}. Note that, for energies exceeding the peak value ($E \gg E_{\mathrm{peak}}$), Hawking emission is exponentially suppressed, and for energies lower than the peak value ($E \ll E_{\mathrm{peak}}$), it falls off as a power law.

For this work, we calculate the spectrum of the emitted particles, $\ud \dot{N}/\ud E$, using the publicly available code \texttt{BlackHawk}\footnote{\url{https://blackhawk.hepforge.org/}} \cite{blackhawk}. We have verified the numerically obtained Hawking emission rates against the semi-analytical emission rates from refs.~\cite{Page:1976df, Page:1976ki}.

Hawking-radiated particles (photons $\upgamma$, electrons e$^-$, positrons e$^+$) interact with the ordinary baryonic matter in the IGM and deposit energy. The energy deposition typically occurs through five different channels which include hydrogen ionization, helium ionization, hydrogen excitation, IGM heating and sub-$\SI{10.2}{\electronvolt}$ continuum photons \cite{darkhistory,HLiu2016}. However, for this work, only hydrogen ionization and IGM heating are relevant because we are ignoring the helium recombination-ionization \cite{Mesinger_11}, hydrogen excitation is also irrelevant for reasons discussed later and sub-$\SI{10.2}{\electronvolt}$ continuum photons are involved in CMB spectral distortions, which is again not important for our work \cite{Slatyer2016_II}. The power density (energy per unit time per unit volume) going into a particular channel `c' is
\begin{multline}
q_{\mathrm{HR,c}}= \int\!\!\int \left[f_{\mathrm{c}}(E_{\upgamma},z)E_{\upgamma}\left(\frac{\ud\dot{N}}{\ud E}\right)_{\upgamma}+2f_{\mathrm{c}}(E_{\ue}-m_{\ue}c^2,z)(E_{\ue}-m_{\ue}c^2)\left(\frac{\ud\dot{N}}{\ud E}\right)_{\ue^{\pm}}\right]\\ \times n_{\mathrm{PBH}}(M_{\mathrm{PBH}})\psi(M_{\mathrm{PBH}})\,\ud M_{\mathrm{PBH}}\,\ud E\,,\label{phr1}
\end{multline}
where $E$ is the total energy of the emitted particle, $n_{\mathrm{PBH}}$ is the number density of PBHs, $\psi(M_{\mathrm{PBH}})$ denotes the probability distribution function of PBH masses, and $f_{\mathrm{c}}(E_{\mathrm{K}},z)$ denotes the ratio of the energy deposited into channel c to the injected energy as a function of kinetic energy of the emitted particle $E_{\mathrm{K}}$ and redshift $z$. We use the numerical data table of $f_{\mathrm{c}}$ corresponding to DM decay from ref.~\cite{darkhistory} which are in close agreement with that of ref.~\cite{HLiu2016}.

In this work, we consider a monochromatic mass distribution of PBHs, for which $\psi(M_{\mathrm{PBH}})$ is a Dirac delta function centred at $M_{\mathrm{PBH}}$. Equation~\eqref{phr1} then simplifies to
\begin{multline}
q_{\mathrm{HR,c}}=\int \left[f_{\mathrm{c}}(E_{\upgamma},z)E_{\upgamma}\left(\frac{\ud\dot{N}}{\ud E}\right)_{\upgamma}+2f_{\mathrm{c}}(E_{\ue}-m_{\ue}c^2,z)(E_{\ue}-m_{\ue}c^2)\left(\frac{\ud\dot{N}}{\ud E}\right)_{\ue^{\pm}}\right]\\\times n_{\mathrm{PBH}}(M_{\mathrm{PBH}})\ud E\,.
\end{multline}
Since PBHs make up the solitary component of the present day DM, the number density of PBHs can be written as
\begin{equation}
n_{\mathrm{PBH}}=f_{\mathrm{PBH}}\frac{\rho_{\mathrm{c}}\Omega_{\mathrm{DM}}}{M_{\mathrm{PBH}}}\,,\label{npbh}
\end{equation}
where $f_{\mathrm{PBH}}$ denotes the fraction of DM composed of PBHs, $\rho_{\mathrm{c}}$ is the critical density of the Universe, and $\Omega_{\mathrm{DM}}$ denotes the present day DM density relative to $\rho_{\mathrm{c}}$. Note that, $q_{\mathrm{HR,c}}$ is almost independent of the redshift because for $M_{\mathrm{PBH}} \geqslant \SI{e15}{\gram}$, the mass loss of PBHs due to Hawking emission is negligible and $f_\mathrm{c}$ has a weak dependence on $z$ for the range considered in this work.
 
Using eq.~\eqref{DEofTk}, the heating term now is 
\begin{multline}
\frac{2q_{\mathrm{HR,heat}}}{3n_{\mathrm{b}}k_\mathrm{B}H}=f_{\mathrm{PBH}}\frac{2m_{\mathrm{b}}}{3k_\mathrm{B}H}\left(\frac{\Omega_{\mathrm{m}}}{\Omega_{\mathrm{b}}}-1\right)\frac{1}{M_{\mathrm{PBH}}}\\\times\int \left[f_{\mathrm{heat}}(E_{\upgamma},z)E_{\upgamma}\left(\frac{\ud\dot{N}}{\ud E}\right)_{\upgamma}+2f_{\mathrm{heat}}(E_{\ue}-m_{\ue}c^2,z)(E_{\ue}-m_{\ue}c^2)\left(\frac{\ud\dot{N}}{\ud E}\right)_{\ue^{\pm}}\right]\ud E\,,\label{qhr}
\end{multline}
where average baryon mass, $m_{\mathrm{b}}$, is given by eq.~\eqref{abm}. Similarly, the ionization rate due to Hawking emission from the ground state of hydrogen atom, required in eq.~\eqref{xe_bulk}, can be written as 
\begin{equation}
\Gamma_{\mathrm{HR}}=\frac{q_{\mathrm{HR,ion}}}{n_{\mathrm{H}}E_{\infty}}\,,
\end{equation}
where $E_{\infty}=\SI{13.6}{\electronvolt}$ is the ionization energy of hydrogen. In figure~\ref{comb} we show the effect on the 21-cm signal when $f_{\mathrm{PBH}}$ is varied for a PBH of mass $M_{\mathrm{PBH}}=\SI{e15}{\gram}$.

The difference between our standard equation of thermal evolution, eq.~\eqref{DEofTk}, and eq.~(11) from previous analysis \cite{Clark} is mainly in accounting of different astrophysical processes. Other than the adiabatic cooling and Compton heating we also consider Ly~$\alpha$ and X-ray heating. On comparing our ionization equations, eq.~\eqref{xe_bulk}, and eq.~(3) from ref.~\cite{Clark} we find the following differences. Firstly, we identify that the collisional ionization is negligible for the epoch of our interest and hence the term with coefficient $\beta$ can be ignored. Secondly, unlike previous literature we incorporate X-ray photons which, other than heating, are also involved in the ionization of IGM. Thirdly, we do not have the Peebles $C$ factor, which can be interpreted as the probability for an atom initially in the first excited state to reach the ground state before being ionized. For redshifts considered in this work, it can be set to 1 to an excellent approximation \cite{Peebles, hyrec}. Lastly, because $C=1$, we can say that the ionizations from the first excited state are negligible. This explains our exclusion of the term $I_{\mathrm{X}_\alpha}(z)$ seen in ref.~\cite{Clark}.

Specific to 21-cm calculation, while ref.~\cite{Clark} used the data from ref.~\cite{Ciardi} to calculate Ly~$\alpha$ coupling, we have taken a more physically motivated approach for this as highlighted in section~\ref{hlc}. Finally, we comment on the differences in PBH modelling. Previous literature (e.g. refs.~\cite{Clark, Halder}) have used the following to estimate the mass loss rate due to Hawking radiation
\begin{equation}
\frac{\ud M_{\mathrm{PBH}}}{\ud t}=-\num{5.34e25}\mathcal{F}(M_{\mathrm{PBH}})\left(\frac{\SI{1}{\gram}}{M_{\mathrm{PBH}}}\right)^2\si{\gram\per\second}\,,\label{mdot}
\end{equation}
where $\mathcal{F}(M_{\mathrm{PBH}})$ is a measure of the number of emitted particle species, normalised to unity for a black hole of mass $M_{\mathrm{PBH}}\gg\SI{e17}{\gram}$ \cite{MacGibbon:1991tj}. Then using the expression of $\dot{M}_{\mathrm{PBH}}$, the rate of energy injection per unit volume was given as $q_{\mathrm{inj}}=-n_{\mathrm{PBH}}\dot{M}_{\mathrm{PBH}}c^2$.

\section{Inference procedure}\label{method}

We now explain the inference procedure that we use to constrain our model parameters from the EDGES data.\footnote{\url{http://loco.lab.asu.edu/edges/edges-data-release/}} We follow a Bayesian procedure with a Gaussian likelihood. Our approach is similar to ref.~\cite{Mirocha_2015}. We emphasise that we use the full information content of the EDGES data by using their measurements at all of 123 redshift points, rather than focusing on some specific features of the signal \cite{Mirocha_2015, Cohen17}.

Let $\Delta T_{\mathrm{b}}^{\mathrm{exp}}=\Delta T_{\mathrm{b}}^{\mathrm{exp}}(z)$ and $\Delta T_{\mathrm{b}}^{\mathrm{theo}}=\Delta T_{\mathrm{b}}^{\mathrm{theo}}(\theta,z)$ represent the data and the model values, respectively, of the 21-cm signal at redshift $z$ and $\theta$ being the set of $n$ parameters that parametrize our model. Our likelihood is then
\begin{equation}
\mathcal{L}(\Delta T_{\mathrm{b}}^{\mathrm{exp}}|\theta)=\prod_{i=1}^{123}\frac{1}{\sqrt{2\pi} \varepsilon_i}\exp\left[-\frac{\left(\Delta T_{\mathrm{b}}^{\mathrm{exp}}-\Delta T_{\mathrm{b}}^{\mathrm{theo}}\right)_i^2}{2\varepsilon_i^2}\right]\,,
\end{equation}
where $\varepsilon_i$ are the 1\,$\sigma$ uncertainties in the data for the redshift bin $i$. The label $i$ in the above equation runs over the 123 data points corresponding to the different redshift bins. We have suppressed $\theta$ and $z$ dependence on the right hand side for simplicity.

We take a constant uncertainty for all the redshift bins, i.e., $\varepsilon_i=\SI{0.05}{\kelvin}$. Some remarks are in order about this. In the result presented by the EDGES collaboration, i.e., in Bowman~et~al.~(2018)~\cite{Bowman}, only the uncertainty on the amplitude of the absorption at a particular redshift of $z\approx17$ is presented. This uncertainty estimate is not useful if one intends to infer constraints using the full frequency range of the EDGES data. This problem has been noted before, and in response other groups have inferred uncertainties for the entire frequency range of the EDGES data. For example, Hills~et~al.~(2018)~\cite{Hills} provide an uncertainty estimate of $\SI{0.025}{\kelvin}$ at all frequencies within the EDGES band. See, e.g., Chatterjee~et~al.~(2021)~\cite{Atrideb}, who use a frequency-independent $\SI{0.025}{\kelvin}$ uncertainty, or Mirocha~\&~Furlanetto~(2019)~\cite{Mirocha_19}, who assumed a frequency-independent $\SI{0.1}{\kelvin}$ uncertainty throughout the EDGES band. Since an important motivation in our work is to use the information from the full EDGES band, we follow the same approach, albeit with a slightly more conservative estimate for the uncertainty of $\SI{0.05}{\kelvin}$. Choosing $\SI{0.025}{\kelvin}$ will not change any of our constraints, except to slow down the MCMC convergence. As we will see below, our constraints on $f_\mathrm{PBH}$ are significantly stronger than other constraints published in the literature.  We stress that this strength is because we have used information from the full EDGES band. Consequently, our inference stays fairly robust against changes to the uncertainty estimates on the EDGES data points. The constraint does not critically depend on the uncertainty on the EDGES absorption profile, but rather on its shape.

\begin{figure}[H]
\centering
\includegraphics[width=\textwidth]{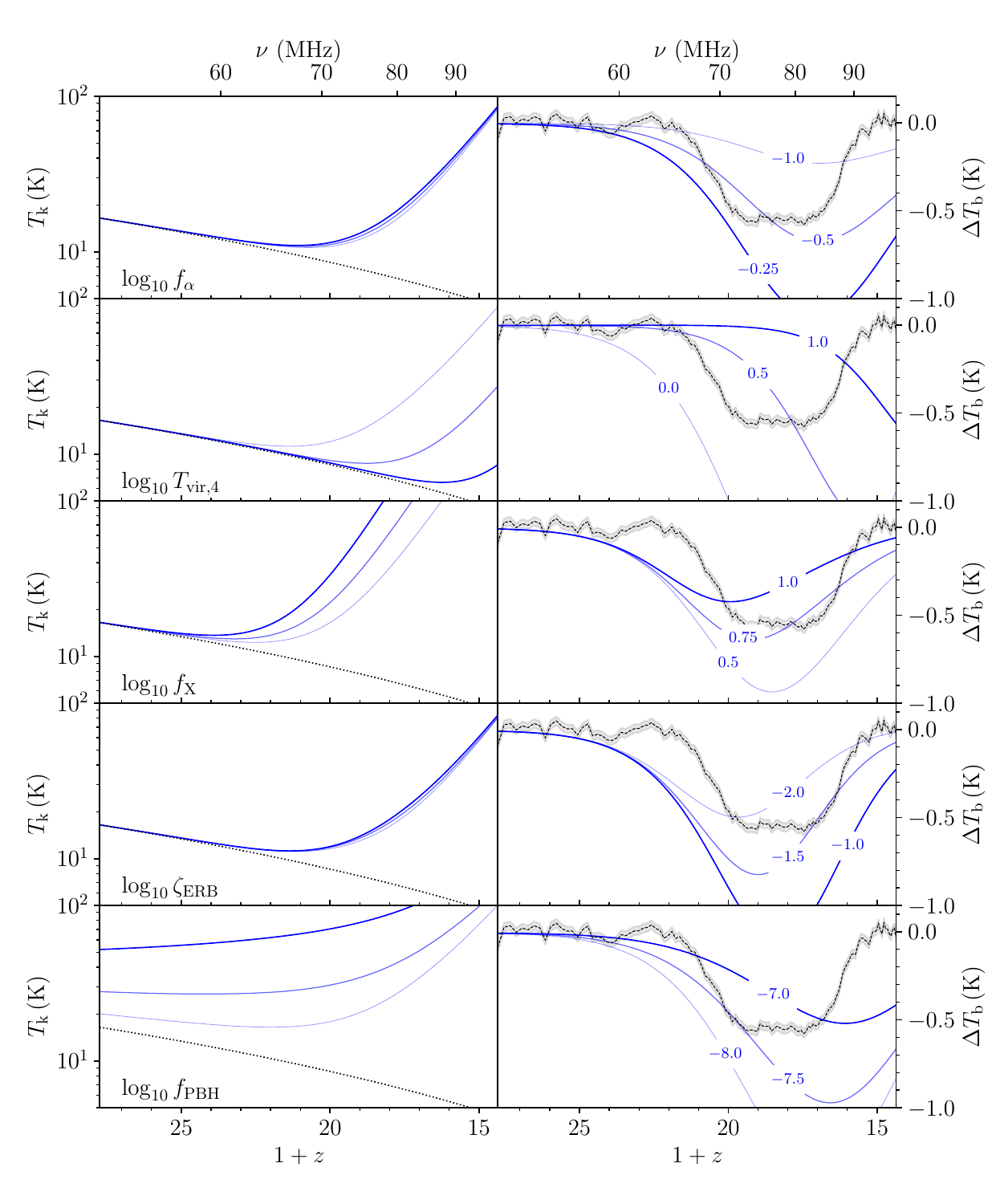}
\caption{Gas temperature evolution (left column) and the corresponding 21-cm signal (right column) for a range of parameter values. For reference, we show the adiabatic thermal evolution in all panels of the left column, and the EDGES measurement of the 21-cm signal in all panels of the right column. In each row, a single parameter is varied, while the remaining parameters are held fixed at $0$, unless one of these fixed parameters is $\log_{10}f_{\mathrm{PBH}}$. When $\log_{10}f_{\mathrm{PBH}}$ is held fixed, it is assigned a value of $-10$, and a PBH mass of $\SI{e15}{\gram}$ is assumed. Because PBH-induced heating has a broadband effect on the signal, even the falling edge of the 21-cm absorption signal can constrain the PBH fraction. See text for a discussion.}
\label{comb}
\end{figure}

We choose uniform priors on our model parameters in the ranges tabulated in table~\ref{Tab1}. This choice of our priors covers a large range of values around the EDGES data, as shown in figure~\ref{comb}. Note that the prior on $\log_{10} f_{\mathrm{PBH}}$ changes with PBH mass. For example, in the presence of X-ray heating the prior required for a PBH of mass $\SI{e15}{\gram}$ is $[-11.0,-9.0]$, while that for a PBH of mass $\SI{e16}{\gram}$ is $[-6.5,-3.0]$.  This choice of a mass-dependent prior might appear odd, but it merely reflects the fact that large values of  $\log_{10} f_{\mathrm{PBH}}$ are obviously ruled out due to the enormous heating rates. One could in principle choose a wide prior, such as $[-11,0]$, for all masses. But this would waste considerable computational effort. For instance, for a $\SI{e15}{\gram}$ black hole, the best-fit $f_{\mathrm{PBH}}$ is $\sim 10^{-9}$, so exploring $f_{\mathrm{PBH}}$ up to 1 for this mass would imply heating rate higher by nine orders of magnitude which is clearly ruled out by the EDGES observation.

\begin{table}[t]
	\centering
	\begin{tabular}{llll}
		\toprule
		Parameter  & Description & Min & Max \\ \midrule
$\log_{10}f_\alpha$ & Controls the strength & &\\ 
& of Ly~$\alpha$ background, eq.~\eqref{epsilon} & $-2$ & 2 \\
$\log_{10}T_{\mathrm{vir,}4}$ & Minimum virial temperature & &\\
& of dark matter haloes, eq.~\eqref{tvir} & $-0.75$ & $1.25$\\
$\log_{10}f_{\mathrm{X}}$ & Controls the strength & &\\
& of X-ray background, eq.~\eqref{epsilonx} & $-1$ & 1\\
$\log_{10}\zeta_{\mathrm{ERB}}$ & Controls the strength & &\\ 
& of excess radio, eq.~\eqref{excess} & $-2$ & 0\\
$\log_{10}f_{\mathrm{PBH}}$ & Fraction of DM & &\\
& in the form of PBHs, eq.~\eqref{npbh} & -- & --\\
		\bottomrule
	\end{tabular}
\caption{Model parameters used in this work, with the ranges of values over which uniform prior PDFs are used. The range of values for $\log_{10}f_{\mathrm{PBH}}$ depends on the mass of PBH under consideration (see text).}\label{Tab1}
\end{table}

\clearpage

We now present a pedagogical discussion of figure~\ref{comb}:
\begin{enumerate}
\item[(a)] Ly~$\alpha$ photons affect the 21-cm signal via two effects: heating and coupling. A higher value of $f_\alpha$ implies a stronger background which in turn means more heating but also a stronger coupling. Heating has a reducing effect on the absorption feature while stronger coupling produces a deeper absorption feature. On the whole the latter wins since Ly~$\alpha$ heating is not a very efficient heating mechanism \cite{Mittal} as evident from $1^{\mathrm{st}}$ panel.

\item[(b)] When the minimum virial temperature is smaller, more star forming haloes are allowed and thus more SFRD. This results in a stronger Ly~$\alpha$ coupling resulting in a deeper 21-cm signal. But the more important role of this parameter is to change the timing of the first drop in the signal without affecting the shape.

\item[(c)] As $f_{\mathrm{X}}$ takes higher values, the X-ray background gets stronger which implies more X-ray heating. This in turn raises $T_{\mathrm{k}}$, and hence $T_{\mathrm{s}}$, thus reducing $\Delta T_{\mathrm{b}}$.

\item[(d)] When we increase $\zeta_{\mathrm{ERB}}$ we allow more excess radio background and hence a stronger contrast between the 21-cm brightness and the background, thus producing a deeper absorption signal.

\item[(e)] Higher $f_{\mathrm{PBH}}$ imply more number of PBHs and hence more heating, which in turn reduces the 21-cm signal. The PBH considered for these plots is of mass $\SI{e15}{\gram}$.
\end{enumerate}

With likelihood and priors ready, we can use the Bayes theorem to construct the posterior distribution as 
\begin{equation}
P(\theta|\Delta T_{\mathrm{b}}^{\mathrm{exp}})\propto \mathcal{L}(\Delta T_{\mathrm{b}}^{\mathrm{exp}}|\theta)\mathcal{P}(\theta)\,,
\end{equation}
where $\mathcal{P}(\theta)$ represents the prior distribution. Since we will use a MCMC implementation \cite{goodman} for sampling $P(\theta|\Delta T_{\mathrm{b}}^{\mathrm{exp}})$, the normalisation of the above is unnecessary. To explore the $n$D parameter space we use the publicly available code \texttt{emcee}\footnote{\url{https://github.com/dfm/emcee}.} \cite{emcee}. We run 64 Markov chains (number of `walkers') and 5000 steps for each parameter, which is a reasonable length since the autocorrelation time for any parameter is not more than $\sim50$. We obtain the initial guess for parameters by maximising $\mathcal{L}(\Delta T_{\mathrm{b}}^{\mathrm{exp}}|\theta)$ when treated as a function of $\theta$. The thermalisation time (number of burn-in steps) is less than 100. Once we have obtained the parameter set, $\theta_{\mathrm{BF}}$, which best explain the model we test the goodness-of-fit by the reduced chi-squared statistics given by $\chi^2/\mathrm{dof}$. The standard definition of $\chi^2$ is given by
\begin{equation}
\chi^2=\sum_{i=1}^{123}\frac{\left[\Delta T_{\mathrm{b}}^{\mathrm{exp}}(z_i)-\Delta T_{\mathrm{b}}^{\mathrm{theo}}(\theta_{\mathrm{BF}},z_i)\right]^2}{\varepsilon_i^2}\,,\label{gof}
\end{equation}
and `dof' stands for degrees of freedom. It is equal to the number of data points minus the number of free parameters employed in the model.

Note that calculating $\Delta T_{\mathrm{b}}^{\mathrm{theo}}$ for any set of parameters during an MCMC simulation can be time consuming and expensive. To overcome this difficulty we prepare our $\Delta T_{\mathrm{b}}^{\mathrm{theo}}$ at some specific grid points in the $n$D space of parameters before running MCMC sampler. With these pre-calculated $\Delta T_{\mathrm{b}}^{\mathrm{theo}}$s we can then estimate $\Delta T_{\mathrm{b}}^{\mathrm{theo}}$ at the desired intermediate parameter set using multi-dimensional linear interpolation. If we have $p$ number of evenly spaced points in each of the $n$ dimensions then we have a total of $p^n$ number of models.

As the X-ray emissivity at Cosmic Dawn is unknown, it is interesting to consider PBHs as the sole heating mechanism that terminates the 21-cm absorption signal. We therefore consider two distinct scenarios for our inference: one without X-ray heating and the other with X-ray heating. In the following subsections we give the details of our analysis for these two cases. The aim in both cases is the same: to obtain the allowed values of the fraction of DM in the form of PBHs, i.e., $f_{\mathrm{PBH}}$ as a function of PBH mass.

\subsubsection*{Case~I: analysis assuming X-ray heating is absent}

In the absence of X-ray heating, the four major physics components affecting the 21-cm signal are SFRD, Ly~$\alpha$ coupling, HR heating and ERB. The degrees of freedom, dof, is $123-4=119$. Our model parameters in this case are 
\begin{equation}
\theta=\{\log_{10} f_\alpha,\; \log_{10} T_{\mathrm{vir,}4},\; \log_{10}\zeta_{\mathrm{ERB}},\; \log_{10} f_{\mathrm{PBH}}\}\,.
\end{equation}
We choose $p=9$ points for each of these parameters, so that we have a total of $9^4=6561$ models explored. However, note that an MCMC simulation with all the parameters varying is not required for each mass. We need to run the 4-parameter MCMC for just one mass, say $M_{\mathrm{PBH}}=\SI{e15}{\gram}$, to obtain the best-fitting parameters. Once this is done, we can fix all parameters other than $f_{\mathrm{PBH}}$, i.e., $T_\mathrm{vir}, f_\alpha$ and $\zeta_{\mathrm{ERB}}$ to their best-fitting values so that for the remaining masses we have only one parameter to vary. We do the analysis for the following PBH masses $$10^{15},\num{2e15},\num{3e15},\ldots\SI{7e16}{\gram}\,,$$which make a total of 16 masses. The reason for this preferred method of analysis is explained as follows. The Hawking emission, $q_{\mathrm{HR,c}}$, has nearly the same trend as a function of time (a constant throughout the epoch of our interest for a given mass) for all masses. Stated differently, $q_{\mathrm{HR,c}}$ for different masses differ by a constant factor (see also the discussion in section~\ref{disc}). Because our parameter is $\log_{10}f_{\mathrm{PBH}}$ rather than $f_{\mathrm{PBH}}$, the above reasoning implies that the probability distribution of $\log_{10}f_{\mathrm{PBH}}$ and its covariances with other parameters will just shift by a certain amount for different masses.

\subsubsection*{Case~II: analysis allowing for X-ray heating}
Four out of five physics components are same as before, but now we have X-ray heating as well. The degrees of freedom, dof, is $123-5=118$. The models parameters are
\begin{equation}
\theta=\{ \log_{10} f_\alpha,\; \log_{10} T_{\mathrm{vir,}4},\; \log_{10} f_{\mathrm{X}},\; \log_{10}\zeta_{\mathrm{ERB}},\; \log_{10} f_{\mathrm{PBH}}\}\,.
\end{equation}
We choose $p=5$ points for each of these parameters, so that we have a total of $5^5=3125$ models explored. Just as with case~I we need not run the 5-parameter MCMC simulation for all masses. Once we have obtained the non-PBH best-fitting parameter values we can set all of them to these, except $f_{\mathrm{PBH}}$, for the remaining masses. We do the analysis for the following PBH masses $$10^{15},\num{2e15},\num{3e15},\ldots\SI{3e17}{\gram}\,,$$which make a total of 21 masses.

Note a very fundamental difference between the two analysis. In case~II we will only obtain an upper bound on $f_{\mathrm{PBH}}$. But case~I produces a detection of PBHs as in this scenario there are no alternative heating sources that can result in the low-redshift rising edge of the 21-cm absorption profile. Case~II is arguably more conservative as some X-ray emission is perhaps easily plausible at these redshifts (see discussion in section~\ref{gastemp}) but until such high-redshift X-ray sources are known to exist, case~I remains a valid possibility. We discuss this further below.

\section{Results}\label{results}
\subsection{Constraints by assuming X-ray heating to be absent}
\begin{figure}
\centering
\includegraphics[width=1\linewidth]{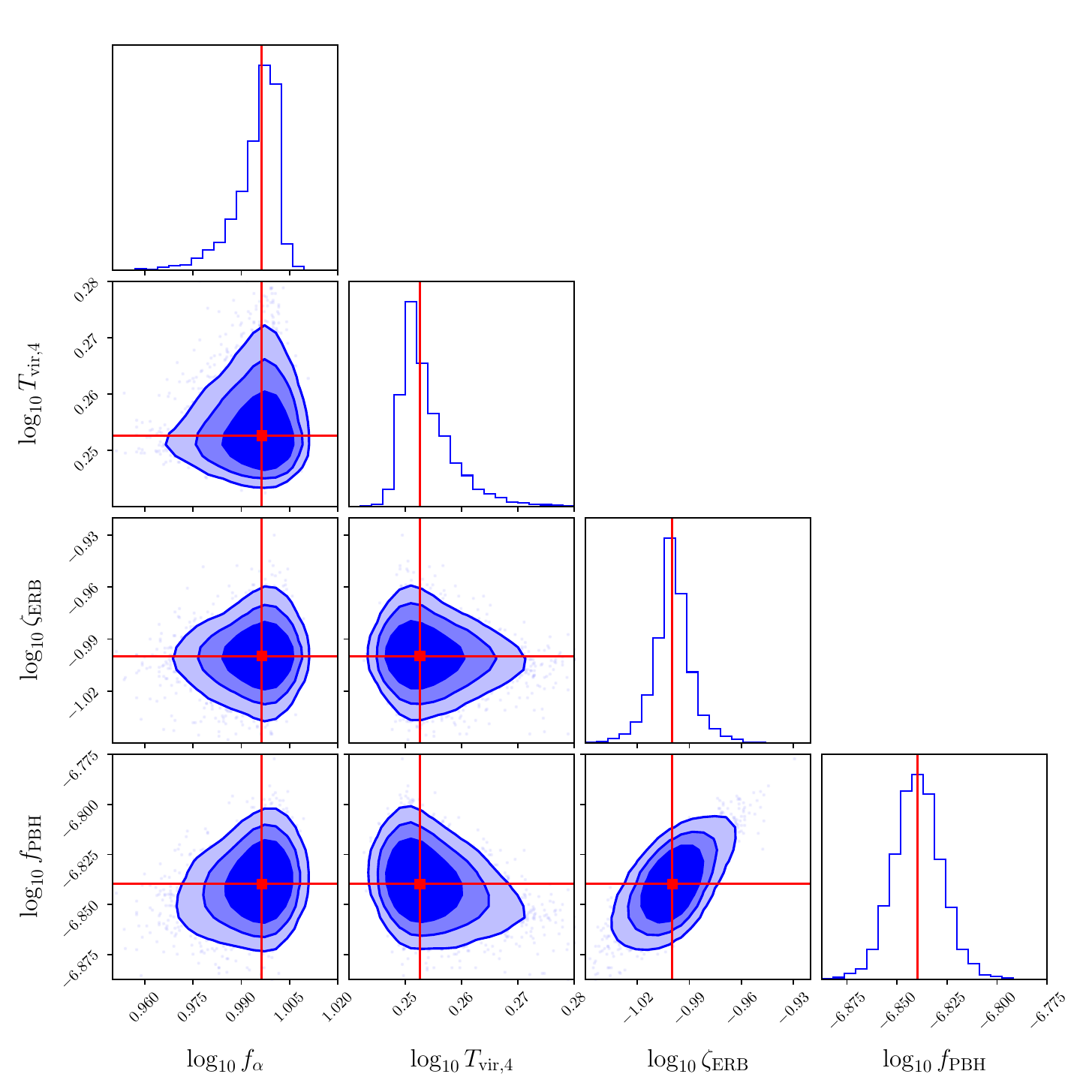}
\caption{Two-dimensional and one-dimensional marginalised posterior probability distributions of parameters for a PBH of mass $M_{\mathrm{PBH}}=\SI{e15}{\gram}$ in a model with no X-ray heating. The contour lines show the 68.3\%, 86.6\% and 95.5\% levels corresponding to 1-sigma, 1.5-sigma and 2-sigma, respectively. The red lines show the median values.}\label{Post1}
\end{figure}

\begin{figure}
\centering
\includegraphics[width=0.8\linewidth]{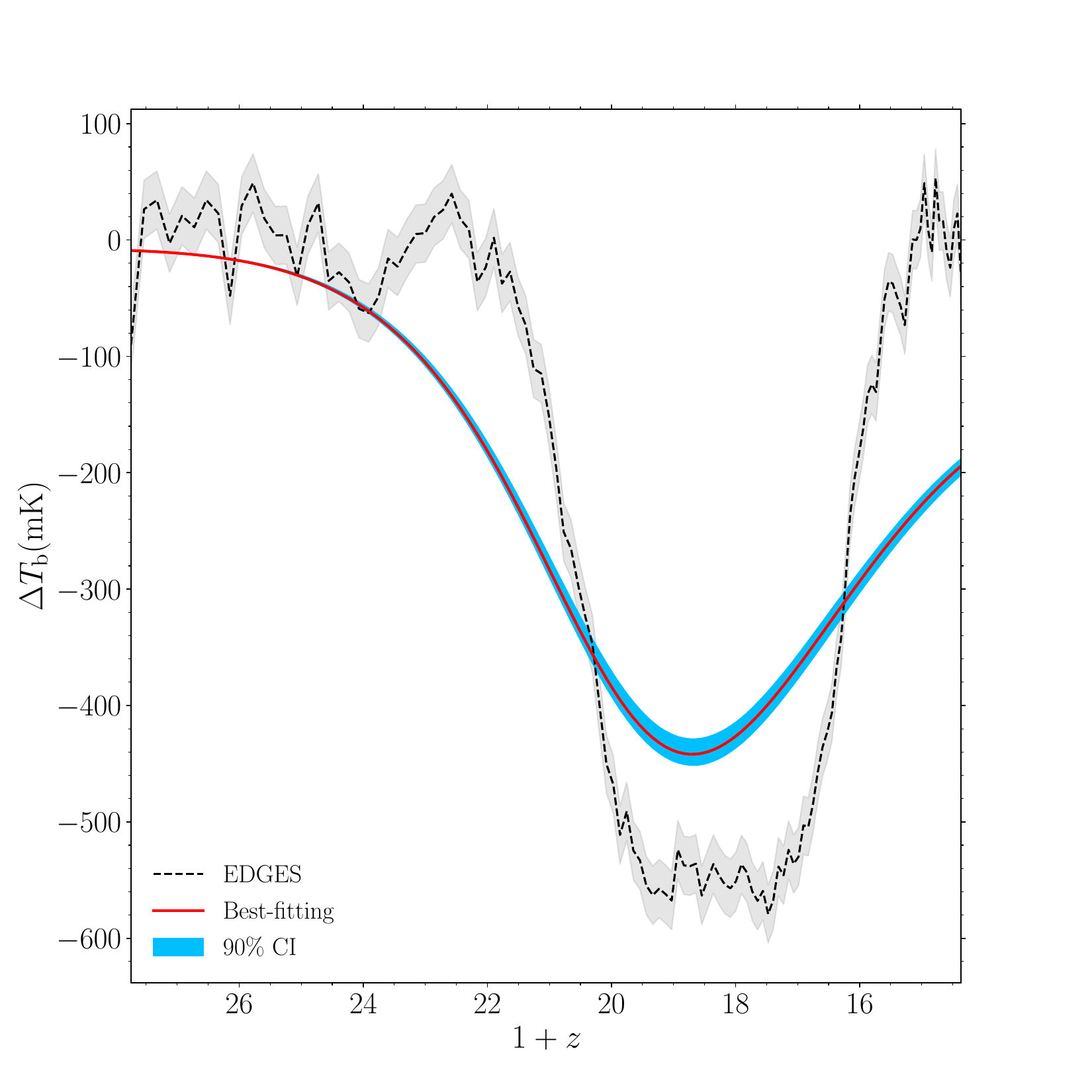}
\caption{Red curve shows the derived posterior median 21-cm signal in the absence of an X-ray background. The blue shaded region shows the 90\% confidence intervals (CI). This is for PBH of mass $\SI{e15}{\gram}$, and corresponds to $\log_{10} f_{\mathrm{PBH}}=-6.8398_{-0.0192}^{+0.0199}$. The EDGES measurement is shown by the black dashed curve, with the grey shaded region around it showing the uncertainty. The goodness-of-fit is $\chi^2/\mathrm{dof}=910.5/119$.}\label{bf1}
\end{figure}

Our marginalised two-dimensional and one-dimensional posterior distributions in the case without X-ray heating are shown in figure~\ref{Post1}. The best-fitting parameter values with 90\% confidence intervals, when we use a PBH of mass $M_{\mathrm{PBH}}=\SI{e15}{\gram}$, are
\begin{align*}
\log_{10} f_\alpha&=0.9964_{-0.0164}^{+0.0057}\,,\\
\log_{10} T_{\mathrm{vir,}4}&=0.2526_{-0.0039}^{+0.0111}\,,\\
\log_{10}\zeta_{\mathrm{ERB}}&=-0.9998_{-0.0187}^{+0.0189}\,,\\
\log_{10} f_{\mathrm{PBH}}&=-6.8398_{-0.0192}^{+0.0199}\,.
\end{align*}
The best-fitting 21-cm signal, 90\% confidence interval and its comparison with the EDGES signal are shown in figure~\ref{bf1} for PBHs of mass $\SI{e15}{\gram}$. Using the definition of goodness-of-fit given in eq.~\eqref{gof} we get $\chi^2/\mathrm{dof}=910.5/119$. For discussion we can divide the complete range into three regions: A, B and C which are the regions on the left of the absorption, the absorption itself and the right side of the absorption, respectively. We see that there is some residual gap in best-fitting curve and data in all A, B and C. Other parameters being fixed, decreasing $f_{\mathrm{PBH}}$ may give the correct absorption depth but will increase the errors in regions A and C. On the other hand if we increase $f_{\mathrm{PBH}}$ we get better fits in A and C but with increased error in B. Thus, we see that there is an optimum value of $f_{\mathrm{PBH}}$ with best fits the signal given some uncertainty $\varepsilon$. A similar reasoning applies for other parameters as well.

\subsection{Constraints allowing for X-ray heating}
\begin{figure}
\centering
\includegraphics[width=0.8\linewidth]{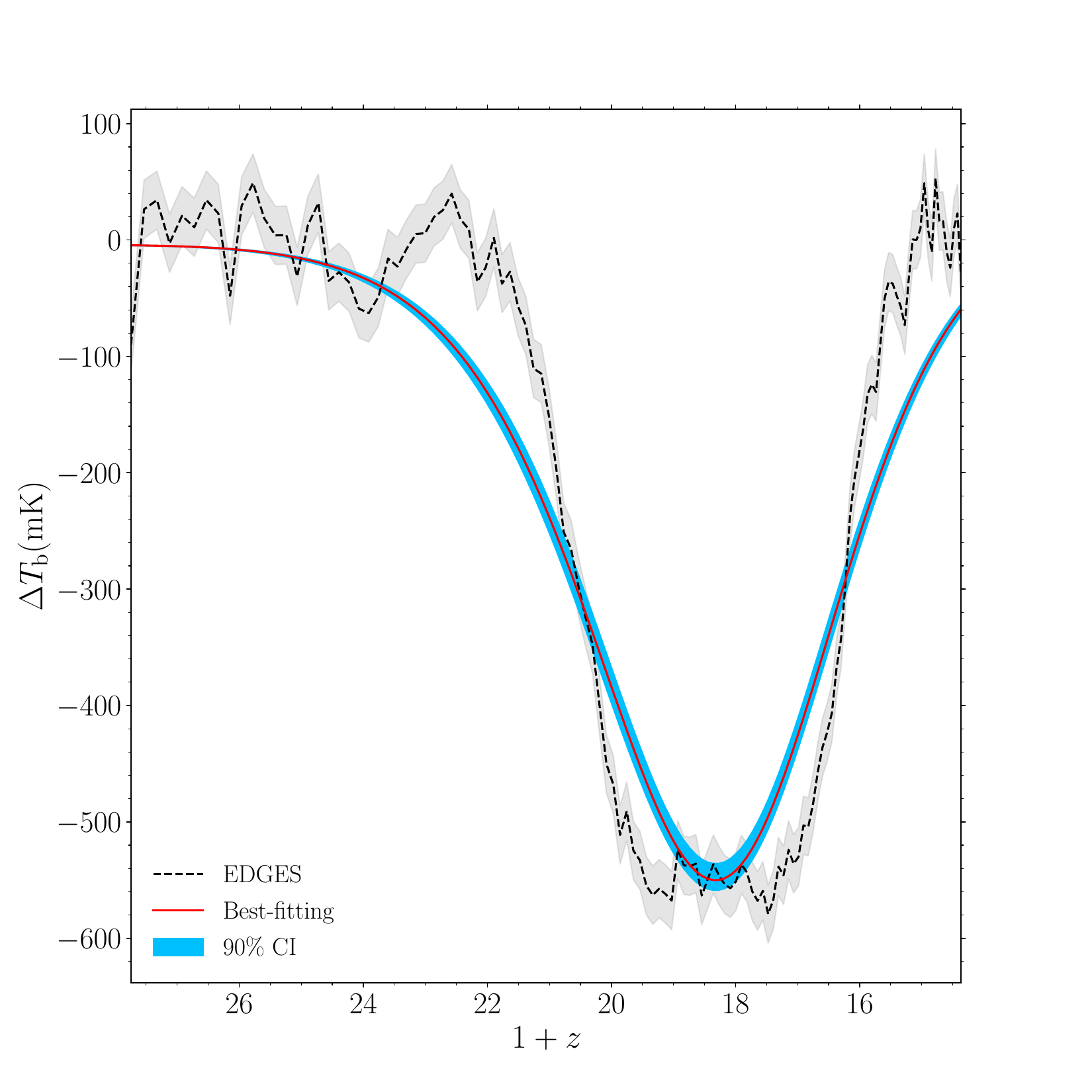}
\caption{Red curve shows the derived posterior median 21-cm signal in the presence of an X-ray background. The blue shaded region shows the 90\% confidence intervals (CI). This is for PBH of mass $\SI{e15}{\gram}$, and corresponds to $f_{\mathrm{PBH}}=10^{-9.73}$ (95\% level). The EDGES measurement is shown by the black dashed curve, with the grey shaded region around it showing the uncertainty. The goodness-of-fit is $\chi^2/\mathrm{dof}=295.3/118$, which is much better compared to case I results.}\label{bf2}
\end{figure}

\begin{figure}
\centering
\includegraphics[width=1\linewidth]{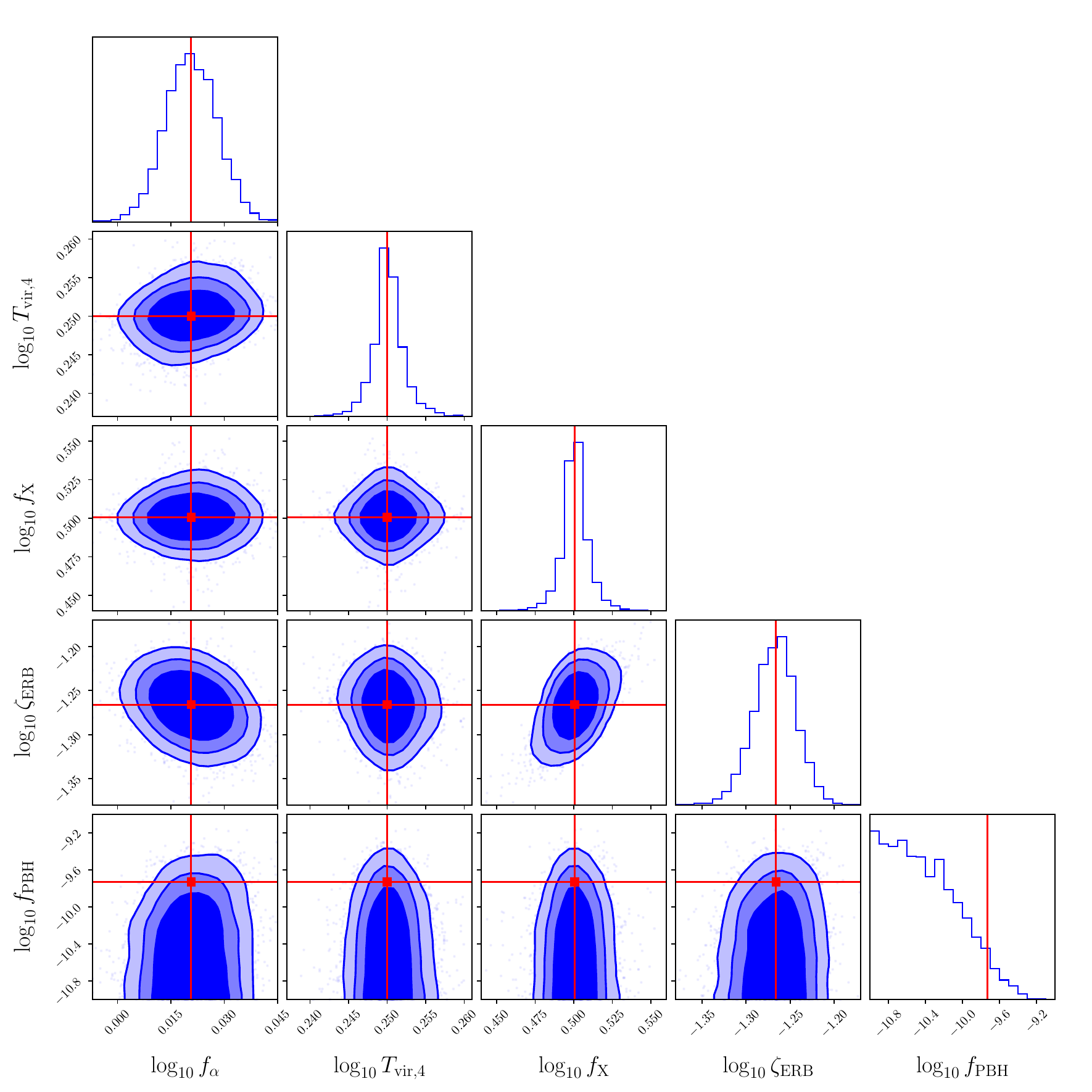}
\caption{Marginalised posterior distributions of parameters for a PBH of mass $M_{\mathrm{PBH}}=\SI{e15}{\gram}$ in the presence of an X-ray background. The contour lines show the 68.3\%, 86.6\% and 95.5\% levels corresponding to 1-sigma, 1.5-sigma and 2-sigma, respectively. The red lines show the median of the probability distribution except in case of $\log_{10}f_{\mathrm{PBH}}$, for which we show the 95\% level.}\label{Post2}
\end{figure}

We now discuss the second case in which we would like to find bounds on $f_{\mathrm{PBH}}$ when there is X-ray heating. The best-fitting non-PBH parameter values obtained are
\begin{align*}
\log_{10} f_\alpha&=0.0207_{-0.0071}^{+0.0073}\,,\\
\log_{10} T_{\mathrm{vir,}4}&=0.2501_{-0.0016}^{+0.0018}\,,\\
\log_{10} f_{\mathrm{X}}&=0.5007_{-0.0069}^{+0.0076}\,,\\
\log_{10}\zeta_{\mathrm{ERB}}&=-1.2666_{-0.0246}^{+0.0224}\,.
\end{align*}
The best-fitting 21-cm signal corresponding to the parameters above, 90\% confidence interval and its comparison with the EDGES signal are shown in figure~\ref{bf2} for PBHs of mass $\SI{e15}{\gram}$. Using the definition of goodness-of-fit given in eq.~\eqref{gof} we get $\chi^2/\mathrm{dof}=295.3/118$. The data thus prefer the model in which X-ray heating accompanies heating due to PBH evaporation; we discuss this point in the next section.  The marginalised two-dimensional and one-dimensional posterior distributions for PBH mass of $M_{\mathrm{PBH}}=\SI{e15}{\gram}$ is shown in figure~\ref{Post2}. We see that in the presence of X-rays, only an upper bound is obtained on $\log_{10}f_{\mathrm{PBH}}$. The model is consistent with the data in the absence of PBHs. We quantify our upper bounds by choosing 95\% levels of the probability distribution of $\log_{10} f_{\mathrm{PBH}}$.

On comparing the best-fitting parameters for the two cases that are only related to setting the depth of the absorption, $f_{\alpha}$ and $\zeta_{\mathrm{ERB}}$, we find that in case~I the values are higher. The Ly~$\alpha$ background required in the absence of X-ray background is nearly 10 times higher in the presence of it. This is understandable because in case II X-ray takes care of the shape which then reduces the requirement of a stronger Ly~$\alpha$ coupling or excess radio background. The minimum virial temperature is roughly the same in both cases because its job is mainly in setting the timing of the first drop in the signal.

\begin{figure}[t]
\centering
\includegraphics[width=0.8\linewidth]{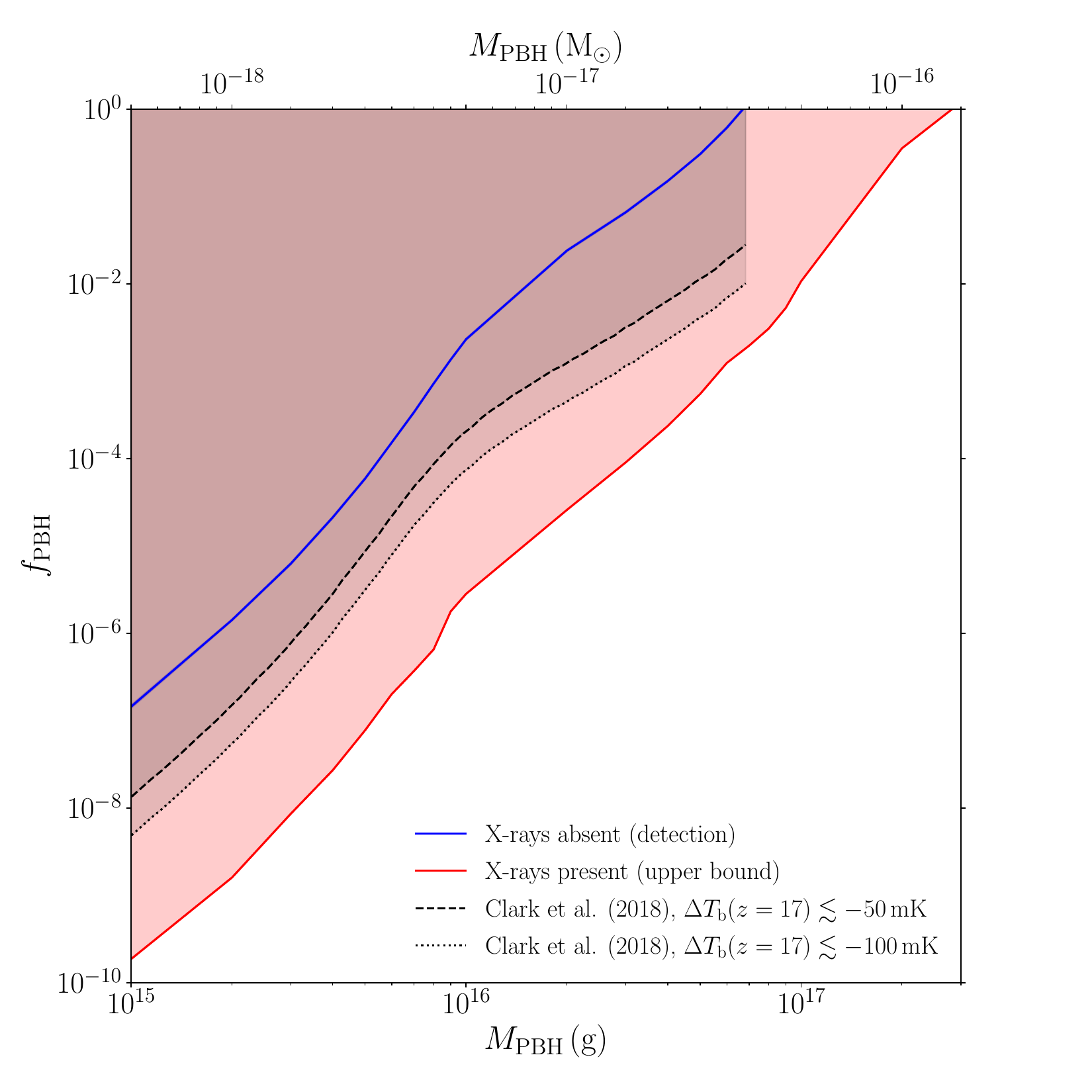}
\caption{Our inferred constraints on the fraction of DM that is in the form of ultralight non-rotating PBHs. In the model with X-rays, the upper limit obtained from this analysis is shown by the red curve; the red shaded region is ruled out. In the model without X-rays, PBHs are the only heating mechanism, so the constraints formally represent a detection of PBHs. This is shown by the blue curve. (The associated uncertainty is too small to be visible on this plot.) For comparison, we also show the result from ref.~\cite{Clark} for $\Delta T_{\mathrm{b}}(z=17)\lesssim\SI{-50}{\milli\kelvin}$ and $\Delta T_{\mathrm{b}}(z=17)\lesssim\SI{-100}{\milli\kelvin}$ by black dashed and black dotted line, respectively.}
\label{main1}
\end{figure}

\begin{figure}
\centering
\includegraphics[width=0.8\linewidth]{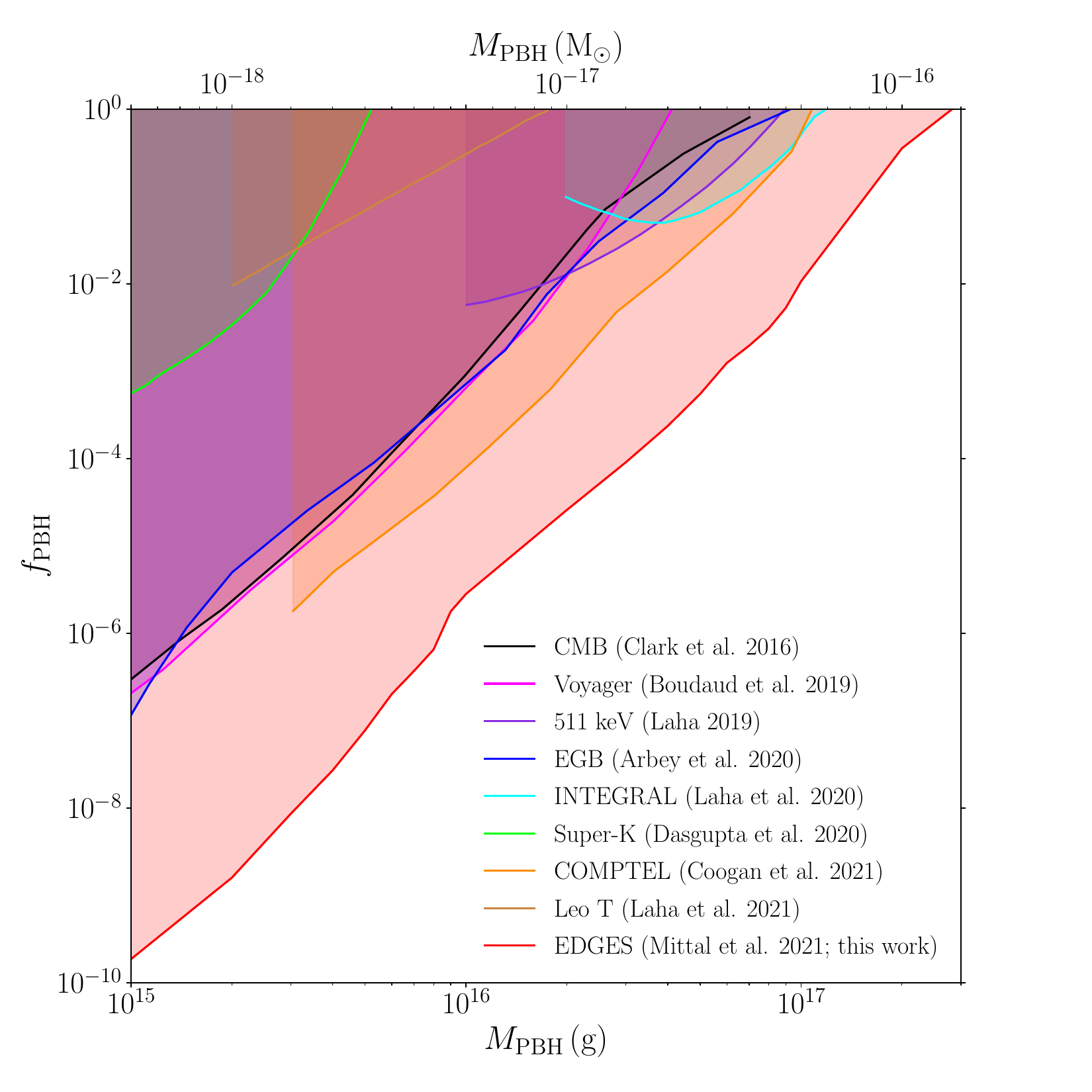}
\caption{Constraints on the fraction of DM that is in the form of ultralight non-rotating PBHs. The exclusion limit obtained from this analysis is shown by the red line; the red shaded region is ruled out. Other exclusion limits, shown for comparison, are from the Planck measurement of CMB (black) \cite{CMBevap}, Voyager measurement of the positron flux (magenta)~\cite{voyager}, SPI/INTEGRAL measurement of the $\SI{511}{\kilo\electronvolt}$ emission line (purple)~\cite{511kev1}, measurement of the EGB (blue) \cite{Arbey:2019vqx}, INTEGRAL (cyan) \& COMPTEL (orange) measurements of the Galactic-centre $\si{\mega\electronvolt}$ gamma-ray flux~\cite{integral,comptel}, diffuse supernovae neutrino background searches at Super-Kamiokande (green) \cite{dasgupta} and Leo~T heating (brown) \cite{Laha:2020vhg}.}\label{main2}
\end{figure}

Figure~\ref{main1} shows our constraints on $f_{\mathrm{PBH}}$ as a function of PBH mass. For comparison we also show the results from previous literature \citep[Clark et al. (2018),][]{Clark} (black dashed and dotted lines) which are somewhat weaker than ours, the reasons for which are discussed in the next section. We see that on log-log scale $f_{\mathrm{PBH}}$ vs $M_{\mathrm{PBH}}$ is approximately a straight line. In absence of X-ray heating we get
\begin{equation}
f_{\mathrm{PBH}}=10^{-6.84}\left(\frac{M_{\mathrm{PBH}}}{\SI{e15}{\gram}}\right)^{3.75}\,,\label{result1}
\end{equation}
with $\sim0.01$ uncertainty (90\%) on $\log_{10}f_{\mathrm{PBH}}$ (for all $M_{\mathrm{PBH}}$) while in presence of X-ray heating
\begin{equation}
f_{\mathrm{PBH}}\leqslant10^{-9.73}\left(\frac{M_{\mathrm{PBH}}}{\SI{e15}{\gram}}\right)^{3.96}\,.\label{result2}
\end{equation}
Figure~\ref{main2} provides a consolidated view of the existing constraints on the fraction of DM composed of ultralight PBHs in the mass range of $\num{e15}$--$\SI{e17}{\gram}$. This includes constraints obtained from the Planck measurement of CMB~\cite{CMBevap,Stocker:2018avm,Poulter:2019ooo,Acharya:2020jbv, Cang_2021}, electron/positron flux measurements by Voyager \cite{voyager}, measurement of the $\SI{511}{\kilo\electronvolt}$ line by SPI/INTEGRAL \cite{511kev1, 511kev2,dasgupta}, measurement of the extra-galactic gamma-ray emission (EGB) by COMPTEL, SMM \cite{BJC,Arbey:2019vqx,Ballesteros:2019exr}, Galactic Centre $\si{\mega\electronvolt}$ gamma-ray measurements by INTEGRAL \& COMPTEL \cite{integral, comptel}, diffuse supernovae neutrino background searches at the Super-Kamiokande neutrino observatory~\cite{dasgupta} and Leo~T heating~\cite{Kim:2020ngi,Laha:2020vhg}. (Appendix~\ref{AppA} lists the numerical values of our inferred constraints.)

\subsection{Constraints on the primordial curvature power spectrum}

If PBHs form due to the collapse of large density perturbations in the very early Universe, exclusion limits on the PBH abundance can be translated to the constraints on the primordial curvature power spectrum~\cite{Bugaev:2008gw,Josan:2009qn,Sato-Polito:2019hws,Kalaja_2019}. This leads to constraints on the primordial curvature power spectrum at small scales that are inaccessible to any other cosmological observable. 

In gravitational collapse a certain fraction of the horizon mass collapses and forms PBHs. If all the PBHs formed at the same epoch, say in a radiation dominated era with a monochromatic mass distribution, we can relate their mass to the present day horizon mass, $M_0=c^3/(2G_{\mathrm{N}}H_0)$, 
\begin{equation}
M_{\mathrm{PBH}} = \gamma\,\sqrt{\Omega_{\mathrm{r}}}\,M_0 \left(\frac{g_0}{g_{\mathrm{i}}}\right)^{1/6}\left(\frac{H_0}{ck}\right)^2\,,\label{mpbh}
\end{equation}
where $\gamma\approx0.2$ is the fraction of collapsed horizon mass \cite{Carr1975, Sato-Polito:2019hws}, $\Omega_{\mathrm{r}}\approx\num{9e-5}$ is the present day radiation density relative to critical density, $k$ denotes scale of horizon re-entry and $g_0=3.38$ $(g_{\mathrm{i}}=106.75)$ denotes the total number relativistic degrees of freedom at present day (at the time of PBH formation) \cite{dof}.

The initial mass fraction $\beta$ of PBHs, is related to the present day PBH fraction of DM ($f_{\mathrm{PBH}}$) through~\cite{Sato-Polito:2019hws}
\begin{equation}
\beta (M_{\mathrm{PBH}}) = f_{\mathrm{PBH}} \left(\frac{g_{\mathrm{i}}}{g_0}\right)^{1/4} \left(\frac{\Omega_{\mathrm{DM}}}{\Omega^{3/4}_\mathrm{r}}\right) \sqrt{\frac{M_{\mathrm{PBH}}}{\gamma M_0}}\,.\label{imf}
\end{equation}
Note that, an assumption made in order to arrive at eqs.~\eqref{mpbh} and \eqref{imf} is that the effective degrees of freedom for entropy are equal to that of energy (cf. ref.~\cite{Kalaja_2019}).

In Press-Schechter theory \cite{Press}, the initial mass fraction of PBHs is equivalent to the probability that the smoothed density field exceeds its threshold $\delta_\mathrm{c}\approx0.42$ \cite{Harada}. Therefore, $\beta$ can also be written as
\begin{equation}
\beta (M_{\mathrm{PBH}}) = 2 \int_{\delta_\mathrm{c}}^{1} \Pi(\delta)\,\ud\delta \approx \mathrm{erfc} \left(\frac{\delta_{\mathrm{c}}}{\sqrt{2}\sigma}\right)\,,\label{variance}
\end{equation}
where $\Pi(\delta)$ denotes the probability density of the density contrast $\delta=\delta(R)$ for a comoving length scale $R$. The probability density is assumed to be a Gaussian of variance $\sigma^2=\sigma^2(R)$ which can be written in terms of curvature power spectrum $\mathcal{P}_\mathcal{R}$ as \cite{Josan:2009qn}
\begin{equation}
\sigma^2=\frac{16}{3}\int_0^{\infty}(kR)^2j^{2}_1\left(\frac{kR}{\sqrt{3}}\right)\ue^{-(kR)^2}\mathcal{P}_\mathcal{R}(k)\frac{\ud k}{k}\,.
\end{equation}
Assuming that the integral in the expression of mass variance $\sigma$ is dominated at $kR\sim1$ we can estimate the curvature power spectrum as
\begin{equation}
\mathcal{P}_\mathcal{R}(k)\approx\frac{3\ue}{16}j^{-2}_1\left(\frac{1}{\sqrt{3}}\right)\sigma^2\,,\label{cps}
\end{equation}
where 
\begin{equation}
j_1(x)=\frac{\sin x-x\cos x}{x^2}\,,
\end{equation}
is the spherical Bessel function.  

\begin{figure}
\centering
\includegraphics[width=0.8\linewidth]{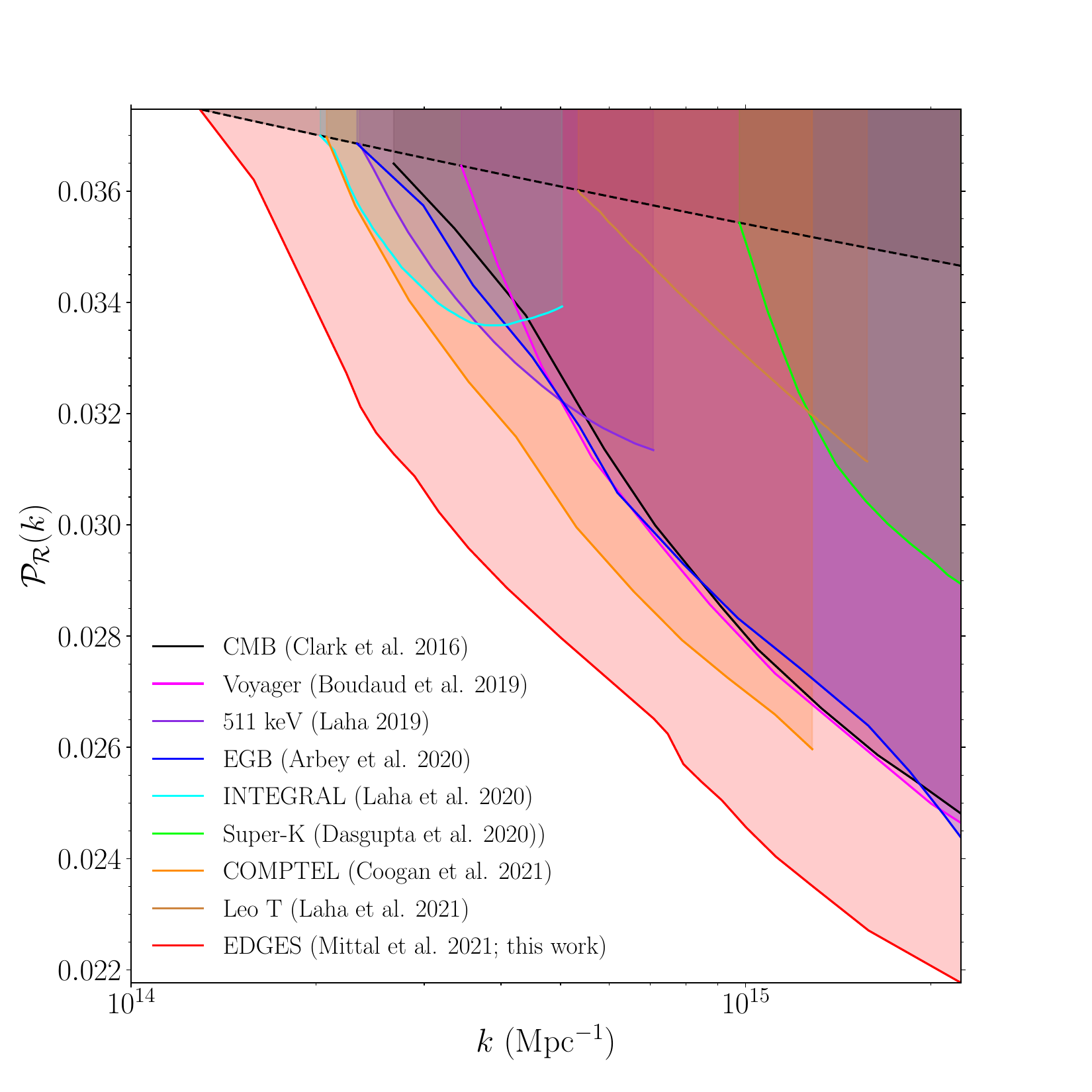}
\caption{Upper limits (red curve) on the curvature power spectrum by translating our upper limits on $f_{\mathrm{PBH}}$ in the presence of X-ray heating. The red shaded region is ruled out.  Constraints translated from the Planck measurement of CMB (black) \cite{CMBevap}, Voyager measurement of the positron flux (magenta)~\cite{voyager}, SPI/INTEGRAL measurement of the $\SI{511}{\kilo\electronvolt}$ emission line (purple)~\cite{511kev1}, measurement of the EGB (blue) \cite{Arbey:2019vqx}, INTEGRAL (cyan) \& COMPTEL (orange) measurements of the Galactic-centre $\si{\mega\electronvolt}$ gamma-ray flux~\cite{integral,comptel}, diffuse supernovae neutrino background searches at Super-Kamiokande (green) \cite{dasgupta} and Leo~T heating (brown) \cite{Laha:2020vhg} are also shown for comparison. The region above the black dashed line is ruled out because in that region the density of PBHs exceeds that of DM i.e. $f_{\mathrm{PBH}}>1$.}\label{figps}
\end{figure}

For a PBH of mass $M_{\mathrm{PBH}}$ we know $f_{\mathrm{PBH}}$, then using eq.~\eqref{imf} we can calculate $\beta$. Using this $\beta$ in eq.~\eqref{variance}, we can find $\sigma$. Using the latter in eq.~\eqref{cps} gives us $\mathcal{P}_\mathcal{R}$. We now have $\mathcal{P}_\mathcal{R}$ as function of $M_{\mathrm{PBH}}$. To get $\mathcal{P}_\mathcal{R}$ as a function of $k$ we finally use eq.~\eqref{mpbh}. Our resulting constraints on the curvature power spectrum corresponding to case~II (X-ray heating included) can be approximated as
\begin{equation}
\mathcal{P}_\mathcal{R}(k)\leqslant \num{2.46e-2}\left(\frac{k}{\SI{e15}{\mega\parsec^{-1}}}\right)^{n_{\mathcal{R}}-1}\,,\label{psfit}
\end{equation}
where $n_{\mathcal{R}}\approx0.806$. Our result along with power spectrum constraints corresponding to other abundance constraints (shown in figure~\ref{main2}) are shown in figure~\ref{figps}. 

\section{Discussion}\label{disc}

\begin{figure}[t]
\centering
\includegraphics[width=0.8\linewidth]{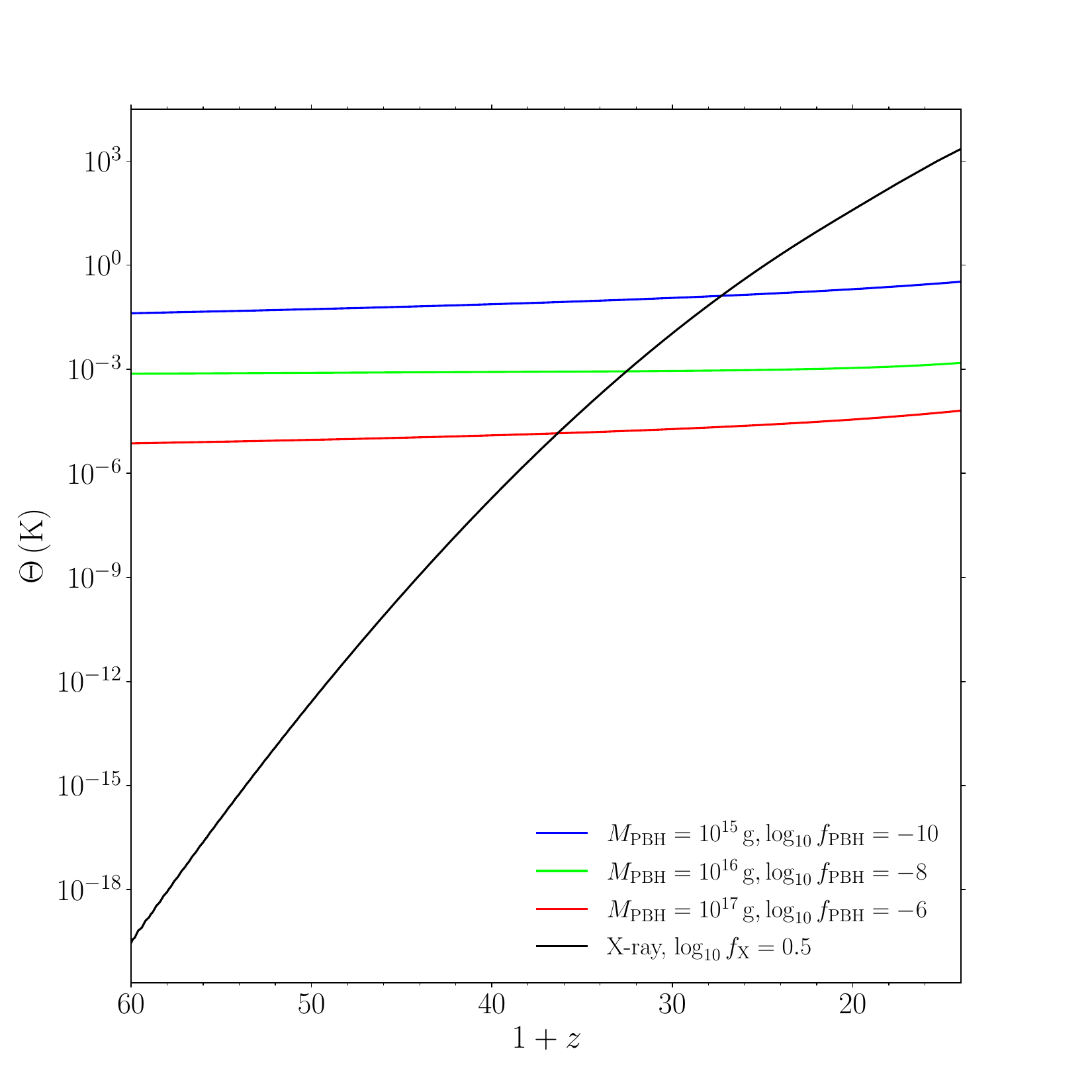}
\caption{A comparison of the heating rate due to PBH with that due to X-rays in our model. The heating rate is expressed here in units of temperature as $\Theta=2q\cdot \left[3n_{\mathrm{b}}k_{\mathrm{B}}H(z)\right]^{-1}$. The blue, green and red solid lines shows HR heating for different combinations of $M_{\mathrm{PBH}}$ and $f_{\mathrm{PBH}}$. The slow rise seen in $\Theta_{\mathrm{HR}}$ is dominantly because of $H(z)$ as $q_{\mathrm{HR,heat}}$ remains constant (since $f_\mathrm{c}(E_{\mathrm{K}},z)$ has a weak dependence on $z$). The black solid line shows X-ray heating for $\log_{10}f_{\mathrm{X}}=0.5$. Because X-ray emission traces halo formation, X-ray heating rises more sharply than PBH-induced heating.}\label{compare}
\end{figure}

Models in which IGM heating is predominantly caused by PBHs predict a qualitatively different thermal history than those in which X-ray heating is dominant. Figure~\ref{compare} shows the contribution to the IGM heating rate by different PBH masses and $f_{\mathrm{PBH}}$. This is shown in comparison with X-ray heating for $\log_{10}f_{\mathrm{X}}=0.5$. X-ray production follows the build-up of dark matter haloes. Consequently, the X-ray heating rate builds up rather rapidly with time. In contrast, PBH heating rate is relatively constant with redshift. The 21-cm absorption signal measured by EDGES has a rapidly rising profile (it changes by $\sim\SI{500}{\milli\kelvin}$ between $z=17$ and 15). PBH-driven models are not capable of reproducing such steep absorption signals because increasing the PBH emissivity at lower redshift also increases it at higher redshifts, thereby weakening the signal itself. Models with X-rays are much better in explaining the steep rising edge of the observed absorption profile. This explains why the best-fitting model in figure~\ref{bf2} is a better fit to the data than the best-fitting model in figure~\ref{bf1}.

This difference between the PBH-induced and X-ray-induced heating rates also helps understand why we only obtain an upper bound on the PBH fraction in the presence of X-rays. The model prefers to explain the rising edge of the absorption feature by means of X-ray emission because enhanced heating by PBHs worsens the signal as a whole due to excess heating at higher redshifts. Stated differently, in the presence of X-ray heating, HR heating does not offer any extra feature that can reduce the tension between the data and the model, which in turn means that all values of $f_{\mathrm{PBH}}$ below a certain maximum are allowed.

The contrast between our inferred constraints, as shown in figure~\ref{main1}, and constraints previously reported in the literature \cite{Clark} illustrate the importance of using the full information content of the data. Constraints reported in ref.~\cite{Clark} were obtained by requiring that the value of the signal $\Delta T_{\mathrm{b}}$ remains below a somewhat arbitrarily chosen threshold of $-50$ or $\SI{-100}{\milli\kelvin}$ at $z=17$, which was taken to be the approximate mid-point of the absorption profile detected by EDGES. In their `standard model', which had no PBH heating, $\Delta T_{\mathrm{b}}(z=17)=\SI{-200}{\milli\kelvin}$. A PBH scenario that led to heating rates that raise $\Delta T_{\mathrm{b}}$ to values greater than $-50$ or $\SI{-100}{\milli\kelvin}$ was deemed to be ruled out by the EDGES data. But the data are richer than this. In general, the three main features in the observed profile are its location, depth and width. If we use all available redshift points from the data, as we do in this work, there are more features to exploit, e.g., the steepness of the rise and fall of the absorption feature. Utilising all of this information is particularly advantageous for heating mechanisms that evolve slowly, such as PBHs. This explains why the constraints from our analysis, shown in figure~\ref{main1}, are tighter than those previously reported in the literature.

Other papers that have done a parameter study include references such as \cite{Cohen17, Mirocha_19, Monsalve_2019, Fialkov_19, Atrideb}. Our choice of parameters is more or less similar to those considered in refs.~\cite{Cohen17, Monsalve_2019, Fialkov_19}. These authors study seven-parameter models with parameters $f_{\star}, V_{\mathrm{c}}, f_{\mathrm{X}}, \nu_{\mathrm{min}}, \alpha, \tau_{\mathrm{e}}$ and $R_{\mathrm{mfp}}$ that represent the star formation efficiency, minimum virial circular velocity of star-forming haloes, X-ray heating efficiency, low frequency cut-off of the X-ray SED, slope of X-ray SED, electron scattering optical depth and mean-free path of ionizing photons, respectively. In this work, we did not treat $f_{\star}$ as a free parameter due to its degeneracy with $f_\alpha$ and $f_{\mathrm{X}}$. The parameter $V_{\mathrm{c}}$ is present in our analysis as $T_{\mathrm{vir}}$ for setting the minimum halo mass for star formation \cite{BL01, DAYAL20181}. The parameter $f_{\mathrm{X}}$ is present in our analysis as well. The other two parameters related to X-ray SED, $\nu_{\mathrm{min}}$, (quantified by energy, $E_0$, instead of frequency) and $\alpha$ (denoted by $w$ in this work) are kept fixed here due to degeneracy with $f_\mathrm{X}$ in the case of $\nu_{\mathrm{min}}$ and weak dependence in the case of $\alpha$, as discussed in section~\ref{fx}. The other two parameters used in the literature, $\tau_{\mathrm{e}}$ and $R_{\mathrm{mfp}}$, are not present in our analysis as we exclusively study the 21-cm signal at Cosmic Dawn, when the effects of reionization are absent. The parameter $f_\alpha$ was not present in previous papers. The advantage of $f_\alpha$ is to scale up Ly~$\alpha$ background without affecting other physics. The minimum $T_{\mathrm{vir}}$ also changes Ly~$\alpha$ background but then it simultaneously affects X-ray background also. The other extra parameter $\zeta_{\mathrm{ERB}}$ is specific to the work that have considered an excess radio (denoted by $\xi$ and $A_{\mathrm{r}}$ in ref.~\cite{Feng_2018} and \cite{Fialkov_19}, respectively).

While this choice of parameters may be conservative, it is justified to ask if other parametrizations are possible. Because little is known about the astrophysics at these redshifts, particularly in the presence of exotic processes such as PBH physics, a wide range of alternative parametrizations can be potentially considered. This `model selection' question is unfortunately out of the scope of this paper due to the possible diversity of models. We leave its study for future work.

\section{Conclusions}\label{Conc}

We inferred constraints on the abundance of uncharged non-rotating primordial black holes (PBHs), assuming a monochromatic distribution in PBH masses, in the mass range $\sim\num{e15}$--$\SI{e17}{\gram}$ using the global 21-cm signal measured by EDGES. Our main conclusions are:
\begin{enumerate}
\item In the absence of X-ray heating, PBH evaporation is the only major heating mechanism in our model. In this scenario, the EDGES measurement formally represents a detection of PBHs. For a $\SI{e15}{\gram}$ PBH we infer a best-fitting value of the fraction of dark matter that is in the form of PBHs as $\log_{10} f_{\mathrm{PBH}}=-6.84\pm0.02$. The fraction $f_{\mathrm{PBH}}$ changes as $\sim M_{\mathrm{PBH}}^{3.75}$ at higher PBH masses (eq.~\ref{result1} and figure~\ref{main1}). However, note that the best-fitting values in this scenario are ruled out by Voyager, EGB and CMB measurements, thus favouring our model with X-ray heating.
\item When X-ray heating is present, we get only an upper bound on the fraction of DM in the form of PBHs. But we find that the data favour this scenario because the X-ray heating rate evolves much more rapidly than the heating rate induced by PBH evaporation. For a $\SI{e15}{\gram}$ PBH we infer $\log_{10} f_{\mathrm{PBH}}\leqslant-9.73$ ($95^{\mathrm{th}}$ percentile). The fraction $f_{\mathrm{PBH}}$ changes as $\sim M_{\mathrm{PBH}}^{3.96}$ towards higher PBH masses (eq.~\ref{result2} and figure~\ref{main1}).  
\item Our constraints on $f_{\mathrm{PBH}}$ are the strongest yet for the PBH mass range of $\sim\num{e15}$--$\SI{e17}{\gram}$. An important reason behind this is that we use the measured 21-cm signal values across the EDGES band. This tightens the limits on PBH evaporation because PBH-induced heating rate evolves very little across the redshift range covered by EDGES.
\item Our inferred values for the non-PBH astrophysical parameters are consistent with observations as well as other analyses. The best-fitting normalisation of Lyman-$\alpha$ emissivity is 1, i.e. $f_{\alpha}\approx1$ which also corresponds to our base model of Population II type stars. Similarly, we have $f_{\mathrm{X}}\approx3$ in which our base model normalisation corresponds to $L_\mathrm{X}$-SFR relation observed for high mass X-ray binaries. The minimum virial temperature required for estimating star formation rate density in Press-Schechter formalism is $T_{\mathrm{vir}}\approx\SI{1.8e4}{\kelvin}$, which is close to the atomic cooling limit. Finally, we require an excess radio background quantified by $\zeta_{\mathrm{ERB}}\approx0.05$, where $\zeta_{\mathrm{ERB}}=1$ corresponds to the maximum observed by ARCADE 2/LWA1.
\item We also derived bounds on the curvature power spectrum at extremely small scales under the assumption of a spherical gravitational collapse based on the Press-Schechter formalism. We get an upper limit of $\mathcal{P_R}=\num{2.46e-2}$ at $k=\SI{e15}{\mega\parsec^{-1}}$, with a $\sim k^{-0.2}$ scaling at other values of $k$ (eq.~\ref{psfit} and figure~\ref{figps}).
\end{enumerate}
This work highlights the usefulness of the global 21-cm signal for probing exotic physical processes. It also shows that global 21-cm measurements contain much more crucially useful information than just the redshift of absorption. The large number of experiments currently underway to probe the 21-cm signal add to the promise of this type of study in future.

\acknowledgments

We thank Vid Ir\v{s}i\v{c}, Ranjan Laha, Hongwan Liu and Akash Kumar Saha for useful discussions. It is a pleasure to acknowledge discussions with members of the REACH collaboration. The work of BD and GK is partly supported by the Department of Atomic Energy (Government of India) research project under Project Identification Number RTI~4002, and by the Max-Planck-Gesellschaft through Max Planck Partner Groups. BD is also supported the Department of Science and Technology (Government of India) through a Swarnajayanti Fellowship.

\bibliographystyle{JHEP}
\bibliography{Biblo}

\newpage

\appendix
\section{Tables of constraints}\label{AppA}

Tables \ref{Tab2} and \ref{Tab3} list the numerical values of our constraints on $\log_{10}f_{\mathrm{PBH}}$ shown in figure~\ref{main1}. These tables are also available in electronic format from the arXiv article web page. 

\begin{table}[h]
\centering
\begin{tabular}{ll}
\toprule
Mass (g)  & $\log_{10}f_{\mathrm{PBH}}$  \\ \midrule
\num{1e15} & $-6.840_{-0.019}^{+0.020}$  \\
\num{2e15} & $-5.848_{-0.012}^{+0.011}$  \\
\num{3e15} & $-5.201_{-0.014}^{+0.014}$  \\
\num{4e15} & $-4.672_{-0.014}^{+0.014}$  \\
\num{5e15} & $-4.227_{-0.014}^{+0.015}$  \\
\num{6e15} & $-3.815_{-0.011}^{+0.011}$  \\
\num{7e15} & $-3.466_{-0.015}^{+0.015}$  \\
\num{8e15} & $-3.141_{-0.013}^{+0.013}$  \\
\num{9e15} & $-2.866_{-0.013}^{+0.012}$  \\
\num{1e16} & $-2.634_{-0.013}^{+0.013}$  \\
\num{2e16} & $-1.619_{-0.012}^{+0.012}$  \\
\num{3e16} & $-1.178_{-0.013}^{+0.013}$  \\
\num{4e16} & $-0.821_{-0.011}^{+0.011}$  \\
\num{5e16} & $-0.513_{-0.010}^{+0.010}$  \\
\num{6e16} & $-0.213_{-0.014}^{+0.014}$  \\
\num{7e16} & $+0.083_{-0.013}^{+0.013}$  \\ \bottomrule
\end{tabular}
\caption{Constraints on $\log_{10}f_{\mathrm{PBH}}$ when no X-ray heating is present. Since PBHs are the only major heating process in this model, the EDGES measurement formally implies a detection. The best-fitting (posterior median) value of $\log_{10}f_{\mathrm{PBH}}$ and the associated 90\% confidence limits are listed. These constraints are shown graphically by the blue line in figure~\ref{main1}.}
\label{Tab2}
\end{table}

\clearpage
\begin{table}[t!]
\centering
\begin{tabular}{ll}
	\toprule
Mass (g)  & $\log_{10}f_{\mathrm{PBH}}$  \\ \midrule
\num{1e15} & $-9.729$  \\
\num{2e15} & $-8.796$  \\
\num{3e15} & $-8.064$  \\
\num{4e15} & $-7.567$  \\
\num{5e15} & $-7.108$  \\
\num{6e15} & $-6.695$  \\
\num{7e15} & $-6.428$  \\
\num{8e15} & $-6.185$  \\
\num{9e15} & $-5.748$  \\
\num{1e16} & $-5.548$  \\
\num{2e16} & $-4.587$  \\
\num{3e16} & $-4.041$  \\
\num{4e16} & $-3.626$  \\
\num{5e16} & $-3.259$  \\
\num{6e16} & $-2.905$  \\
\num{7e16} & $-2.706$  \\
\num{8e16} & $-2.512$  \\
\num{9e16} & $-2.275$  \\
\num{1e17} & $-1.970$  \\
\num{2e17} & $-0.448$  \\
\num{3e17} & $+0.080$  \\
\bottomrule
\end{tabular}
\caption{Constraints on $\log_{10}f_{\mathrm{PBH}}$ in the presence of X-ray heating. Only an upper limit on $\log_{10}f_{\mathrm{PBH}}$ is obtained in this model; we list the $95^{\mathrm{th}}$ percentile here. These constraints are shown graphically by the red line and the shaded region in figure~\ref{main1}.}\label{Tab3}
\end{table}
\vspace*{\fill}

\end{document}